\documentclass[lettersize,journal]{IEEEtran}

\usepackage{amsmath,amsfonts}
\usepackage{algorithmic}
\usepackage{algorithm}
\usepackage{array}
\usepackage[caption=false,font=footnotesize,labelfont=rm,textfont=rm]{subfig}
\usepackage{textcomp}
\usepackage{stfloats}
\usepackage{url}
\usepackage{verbatim}
\usepackage{graphicx}
\usepackage{cite}
\usepackage{siunitx}
\usepackage{booktabs}
\usepackage{svg} 
\usepackage{pifont}
\usepackage{url}
\usepackage{ragged2e} 
\usepackage{booktabs,makecell, multirow, tabularx}
\usepackage{dsfont}
\usepackage{algorithm}
\usepackage{algorithmic}

\begin{document}

\title{Environment Sensing-aided Beam Prediction with Transfer Learning for Smart Factory}

\author{Yuan Feng, Chuanbing Zhao, Feifei Gao, Yong Zhang, and Shaodan Ma
	\thanks{
		Y. Feng, C. Zhao, and F. Gao are with the Department of Automation, Tsinghua University, State Key Lab of Intelligent Technologies and Systems, Tsinghua University, State Key for Information Science and Technology (TNList), Beijing 100084, P. R. China (e-mail: feng-y22@mails.tsinghua.edu.cn; zcb23@mails.tsinghua.edu.cn; feifeigao@ieee.org).
	}
	\thanks{Y. Zhang is with School of Electronic and Information Engineering, 
		Beijing Jiaotong University, 100044, P. R. China (e-mail: zhangy@bjtu.edu.cn). 
	}
	\thanks{S. Ma is with the State Key Laboratory of Internet of Things for Smart City and the Department of Electrical and Computer Engineering, University of Macau, Macao SAR, China (e-mail:
		shaodanma@um.edu.mo).
	}

	}
\maketitle
\thispagestyle{empty}

\maketitle
\vspace{-15mm}
\begin{abstract}
In this paper, we propose an environment sensing-aided beam prediction model for smart factory that can be transferred from given environments to a new environment. In particular, we first design a pre-training model that predicts the optimal beam by sensing the present environmental information. When encountering a new environment, it generally requires collecting a large amount of new training data to retrain the model, whose cost severely impedes the application of the designed pre-training model. Therefore, we next design a transfer learning strategy that fine-tunes the pre-trained model by limited labeled data of the new environment. Simulation results show that when the pre-trained model is fine-tuned by 30\% of labeled data from the new environment, the Top-10 beam prediction accuracy reaches 94\%. Moreover, compared with the way to completely re-training the prediction model, the amount of training data and the time cost of the proposed transfer learning strategy reduce 70\% and 75\% respectively.
\end{abstract}

\begin{IEEEkeywords}
Environment sensing, mmWave, transfer learning, beam prediction.
\end{IEEEkeywords}

\section{Introduction}

\IEEEPARstart{T}{he} fourth generation industrial revolution (Industry 4.0) integrates artificial intelligence, the Internet of Things, big data analysis, and automation technologies into manufacturing and production, which requires vast amounts of real-time data transmission~\cite{s22072688,meindl2021four,lasi2014industry}. The coming 6G system, as a key enabling technology for Industry 4.0, is urgently seeking for new technological breakthroughs, particularly in exploring entirely new resource dimensions~\cite{9145564,9390169,LU2020100158}. 
Since the wireless channel, characterized by the propagation of electromagnetic waves through reflection, scattering, refraction, and multipath effects, is intricately linked to the communications environment~\cite{andersen1995propagation}, exploiting out-of-band information from environmental sensing presents a novel resource dimension and could benefit channel-related tasks like beam prediction, codebook design, blockage prediction, and received signal power prediction~\cite{9529181, 9771564,  8642397,marasinghe2021lidar, 10412143, alrabeiah2020millimeter, xu2022computer, chen2022computer,  nishio2019proactive, shi2015convolutional, charan2021vision, yang2023environment}.

Specifically, the authors of~\cite{9529181} extract environmental features from sub-6 GHz channels to predict the beam direction of high-frequency millimeter waves, which greatly reduces the downlink training overhead of mmWave communications. In~\cite{9771564}, the authors use millimeter wave radar to sense user's locations in a real-world traffic scenario and design a beam prediction network based on deep learning. Experiment results show that the Top-5 beam prediction accuracy is over 90\% which saves 93\% of the beam training overhead.
The authors of~\cite{8642397} design a deep neural network~(DNN) with LIDAR data as input to detect the line-of-sight~(LOS) path of communications links. Then the detection results are used to assist base station~(BS) to select the optimal beam. When there is a LOS path, the Top-10 beam prediction accuracy reaches 90\%, whereas when there is no LOS path, the Top-10 beam prediction accuracy is only 50\%.
In~\cite{marasinghe2021lidar}, the point cloud of indoor environment is collected by the LIDAR, and then a long short-term memory~(LSTM) network is used to predict the movement trajectory of people to determine blockage.  
In addition, since visual images contain rich environmental information, some scholars use environmental images to assist in completing channel related tasks.
The authors of~\cite{10412143} extract  communications-related features from environmental images in a smart factory to predict blockage, reference signal receiving power (RSRP), and beam direction.
The authors of~\cite{alrabeiah2020millimeter} utilize a camera installed on BS to capture images of the communications environment. Then, the environmental images and the sub-6 GHz channels are both input into a neural network (NN) to predict millimeter-wave beam and blockage.
In~\cite{xu2022computer}, the authors leverage images taken by user cameras to extract the size and the location information of the dynamic vehicles, and design a DNN to infer the optimal beam pair for transceivers without any pilot signal overhead.
With the assistance of environmental images from different perspective, the authors of~\cite{chen2022computer} only need a simple snapshot of the environment to design a codebook for a specific site, which saves the cost of time, human resources, as well as hardware installation.
In~\cite{nishio2019proactive}, the authors install access points~(APs) and stations on both sides of the indoor corridor, and use a depth camera to capture the depth images of communications environment. The convolutional long short-term memory~(ConvLSTM) network~\cite{shi2015convolutional} is used to extract depth features and predict the received signal power when pedestrians pass through the corridor, thereby assessing whether the link is blocked. 
The authors of~\cite{charan2021vision} utilize visual data captured by red-green-blue (RGB) cameras at the BSs to predict incoming blockages. Then, the prediction results are used by the wireless network to proactively initiate hand-off decisions and to avoid any unnecessary latency. In~\cite{yang2023environment}, semantic information is extracted from environmental image data, selectively encoded based on its task-relevance, and then fused to make decisions for channel related tasks. This scheme can reduce system overheads such as storage space and computational cost while achieving satisfactory prediction accuracy and protecting user privacy.

From \cite{9529181, 9771564,  8642397,marasinghe2021lidar, 10412143, alrabeiah2020millimeter, xu2022computer, chen2022computer,  nishio2019proactive, shi2015convolutional, charan2021vision, yang2023environment}, it is seen that environment sensing-aided wireless communications schemes does not require accurate channel models or any pilot overhead. 
However, the existing works all focus on designing models for a specific environment, and every time when facing a new environment, they need to re-collect the entire dataset and retrain the model, whose cost significantly hinders the real-world application. 
Hence, there is a pressing need to develop a new scheme that can efficiently adapt to a new environment with a small amount of data.

In this paper, we propose an environment sensing-aided beam prediction model that can be transferred from the given environments to a new environment with limited labeled data. 
Considering that smart factory is a typical scenario in Industry 4.0~\cite{cheffena2016industrial}, we take it as a case study.
The main contributions of this paper are as follows:
\begin{itemize}
	\item We first design an environment-sensing framework for dynamic scatterers detection and static environment reconstruction. Dynamic scatterers are separated from continuous frame semantic segmentation maps, while dynamic features are extracted by 2D convolutional neural network~(CNN).
	Static environments are characterized by pseudo image sequence and static features are extracted by 3D CNN. Then, we propose a pre-training model that predicts the optimal beam with dynamic and static features.
%
	\item We design a transfer learning methodology for the proposed beam prediction model, where 
	the pre-training model is trained using the dataset from given environments, while the pre-trained model is fine-tuned using the dataset from a new environment. Then, the proposed pre-training model is transferred from given environments to a new environment.
	
	\item Simulation results show that the Top-5 beam prediction accuracy of pre-trained model reaches 97\% in given environments. When the pre-trained model is fine-tuned by 30\% of labeled data of the new environment, the Top-10 beam prediction accuracy reaches 94\%. Compared with completely re-training the prediction model, the amount of training data and the time cost of the proposed transfer learning strategy reduce 70\% and 75\% respectively.

\end{itemize}

The rest of this paper is organized as follows. Section \ref{system model} describes the system model. Section \ref{vision-aided} presents the design of the environment sensing framework for dynamic scatterers detection and static environments reconstruction. Section \ref{trans} proposes the beam prediction pre-training model and transfer learning strategy. Section \ref{simulation} shows the performance of the proposed scheme and discusses the simulation results. Finally, Section \ref{conclusion} draws the conclusions.

Notation: $\boldsymbol{A}$ is a matrix or tensor; $\boldsymbol{\mathcal{A}}$ is a set; $\boldsymbol{a}$ is a vector; $a_i$ is the $i$-th element of $\boldsymbol{a}$; $a$ is a scalar; $a_{i}^{j}$ is the element of the $i$-th column and the $j$-th row in $\boldsymbol{A}$; $\boldsymbol{a}_i$ is the $i$-th column of $\boldsymbol{A}$; $\mathcal{CN}(\boldsymbol{m}_g, \boldsymbol{R}_g)$ is the complex Gaussian random distribution with mean $\boldsymbol{m}_g$ and covariance $\boldsymbol{R}_g$.

\section{System Model \label{system model}}
\begin{figure}[!t]
	\centering
	\includegraphics[width=3in]{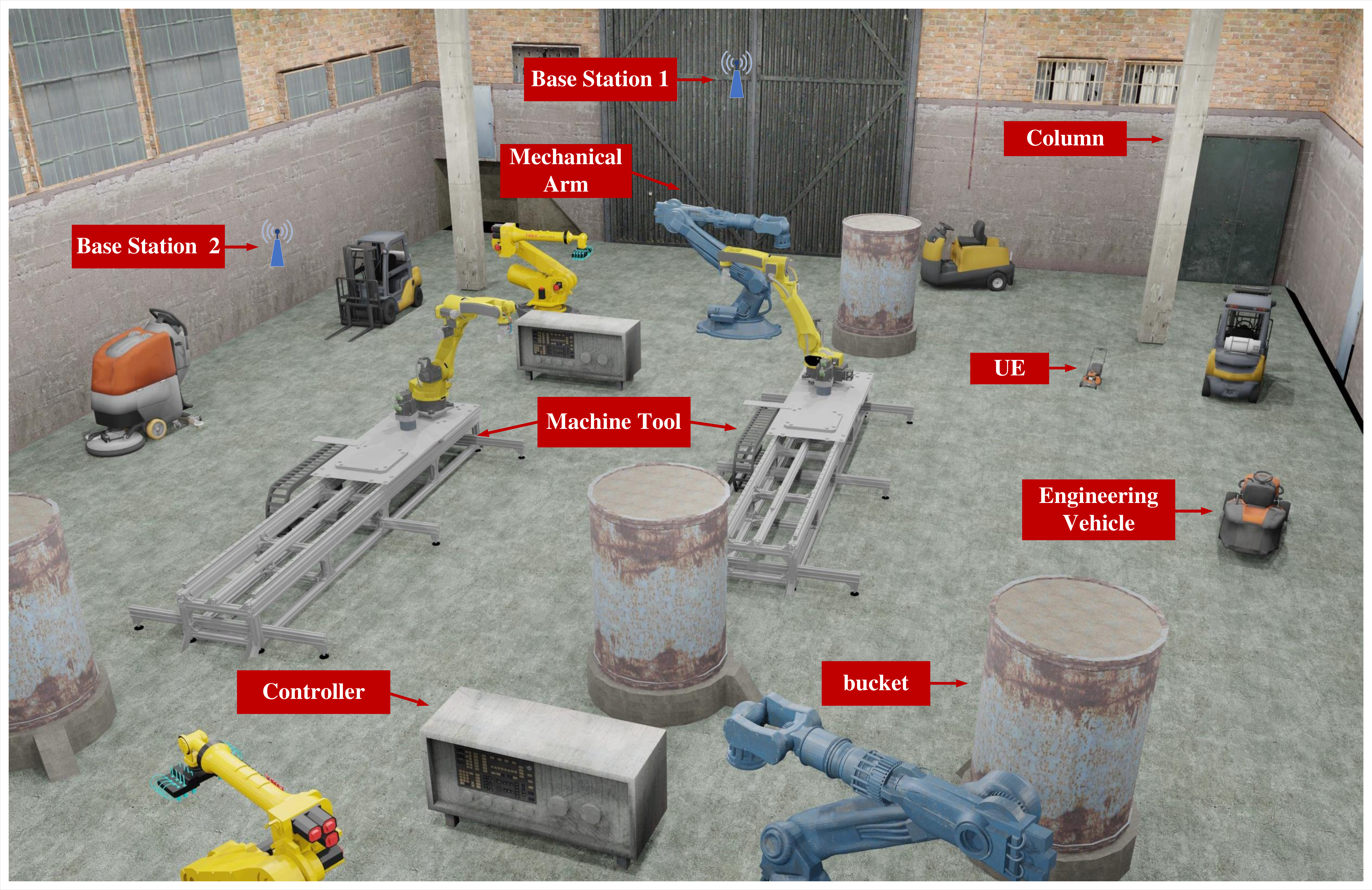}
	\caption{Smart factory scene and equipment such as UE, engineering vehicles, mechanical arms, machine tools, controllers, etc.}
	\label{fig1}
\end{figure}
Let us consider a downlink orthogonal frequency-division multiplexing (OFDM) mmWave communications system in a smart factory, as shown in Fig.~\ref{fig1}. The length, width, and height of the factory space are $l$ m, $w$ m, and $h$ m respectively. There are many types of industrial equipment in the factory like machine tools, mechanical arms, engineering vehicles, steel crossbeams, and steel columns. Multiple downward-looking cameras and a LIDAR are deployed on the ceiling to capture environmental images and 3D point clouds. Moreover, multiple BSs equipped with a uniform planar array (UPA) of $N_t=N_t^x \times N_t^y $ antennas are mounted at a height of $h_B$ m. The user equipment (UE) could be any engineering vehicle in the factory and is equipped with a UPA of $N_r=N_r^x \times N_r^y $ antennas. Considering the cost and the practicality, the BSs and the UE are assumed to have only one radio frequency chain. The downlink received signal at the UE for the $k$-th subcarrier can be expressed as
\begin{equation}
	\label{equ1}
	y_k = {\boldsymbol{w}}_{r,i}^H {\boldsymbol{H}}_{k, i} {\boldsymbol{w}}_{t,i} s_k + {\boldsymbol{w}}_{r,i}^H {\boldsymbol{n}_k},
\end{equation}
where $ {\boldsymbol{H}}_{k, i} \in \mathbb{C}^{N_r \times N_t} $ is the downlink channel matrix at the $k$-th subcarrier between the $i$-th BS and the UE, $ y_k \in \mathbb{C} $ and $s_k \in \mathbb{C}$ are the received and the transmitted signals respectively, ${\boldsymbol{w}}_{r,i} \in \mathbb{C}^{N_r \times 1}$ and ${\boldsymbol{w}}_{t,i} \in \mathbb{C}^{N_t \times 1}$ are the received beam selection vector and the transmitted beam selection vector respectively, and $ {\boldsymbol{n}_k} \in \mathbb{C}^{N_r \times 1} $ is the Gaussian noise vector at the $k$-th subcarrier with $ {\boldsymbol{n}_k} \sim \mathcal{CN}(0,\sigma^2 \boldsymbol{I}_{N_r}) $. The transmitted signal $s_k$ satisfies ${\rm E}\{|s_k|^2\}=P_k$.
According to the widely used geometric channel model~\cite{ali2017millimeter}, the channel matrix ${\boldsymbol{H}_k}$ can be expressed as
\begin{equation}
	\label{equ2}
	{\boldsymbol{H}_k} = \sum_{n=0}^{N-1} \sum_{r=1}^{R} \alpha_r e^{-j \frac{2\pi k}{K}n} g_T(nT_s-\tau_r) \mathbf{a}_r(\phi_r^A,\theta_r^A) \mathbf{a}_t^H(\phi_r^D,\theta_r^D),
\end{equation}
where $R$ is the number of signal propagation paths between each pair of the transmitter and the receiver; $ \alpha_r = \sqrt{P_r} e^{j\Phi_r}$ is complex path gain of the $r$-th path; $P_r$ and $\Phi_r$ are the received signal power and the received signal phase of the $r$-th path respectively; $K$ is the number of subcarriers; $g_T(nT_s-\tau_r) $ is the shaping pulse that is usually set as the raised cosine function with roll off coefficient 0.1; $T_s$ represents the sampling time while $N$ denotes the cyclic prefix length (assuming that the maximum delay is less than $NT_s$), and $\tau_r$ is the time delay of the $r$-th path; $\phi_r^A$ and $\theta_r^A$ are the azimuth and the elevation angles of arrival (AOA) for the $r$-th path; $\phi_r^D$ and $\theta_r^D$ are the azimuth and the elevation angles of departure (AOD) corresponding to the $r$-th path;
$ \mathbf{a}_r(\phi_r^A,\theta_r^A)\in \mathbb{C}^{N_r \times 1}$ and $  \mathbf{a}_t(\phi_r^D,\theta_r^D) \in \mathbb{C}^{N_t \times 1}$  are the corresponding steering vectors given by
\begin{equation}
	\label{equ3}
	\mathbf{a}_r(\phi_r^A,\theta_r^A)=\mathbf{a}_r^y(\phi_r^A,\theta_r^A) \otimes \mathbf{a}_r^x(\phi_r^A,\theta_r^A),
\end{equation}
where $\mathbf{a}_r^x(\cdot)$, $\mathbf{a}_r^y(\cdot)$ are the steering vectors in the $x$, $y$ directions respectively, and $ \otimes $ represents Kronecker product. 
Moreover, $\mathbf{a}_r^x(\cdot)$, $\mathbf{a}_t^y(\cdot)$ can be expressed as
\begin{equation}
	\label{equ4}
	\begin{split}
		\mathbf{a}_r^x(\phi_r^A,\theta_r^A) &= [1,e^{j\pi\sin{\theta_r^A}\cos{\phi_r^A}}, ... , e^{j\pi (N_r^x-1)\sin{\theta_r^A}\cos{\phi_r^A}}]^T,\\
		\mathbf{a}_r^y(\phi_r^A,\theta_r^A) &= [1,e^{j\pi\sin{\theta_r^A}\sin{\phi_r^A}}, ... , e^{j\pi (N_r^y-1)\sin{\theta_r^A}\sin{\phi_r^A}}]^T,\\
	\end{split}
\end{equation}
where $N_r^x, N_r^y$ are the numbers of the received antennas in the $x$, $y$ directions and $N_r=N_r^x N_r^y$. The steering vectors for the AOD are defined similarly. The optimal beam pair $ ( \boldsymbol{w}_r^{opt}, \boldsymbol{w}_t^{opt} )$ will be  selected from the transmit beam codebook $\boldsymbol{\mathcal{W}}_r = \{\boldsymbol{w}_{r,1}, \boldsymbol{w}_{r,2}, ..., \boldsymbol{w}_{r,N_r}\}$ and the receive beam codebook  $\boldsymbol{\mathcal{W}}_t = \{\boldsymbol{w}_{t,1}, \boldsymbol{w}_{t,2}, ..., \boldsymbol{w}_{r,N_t}\}$ by maximizing the transmission rate, i.e.,
\begin{equation}
	 ( \boldsymbol{w}_r^{opt}, \boldsymbol{w}_t^{opt} ) = \arg\max \frac{1}{K} \sum_{k=1}^{K} \log_2 (1+\frac{P_k}{\sigma^2}|\boldsymbol{w}_{r}^H\boldsymbol{H}_k\boldsymbol{w}_{t}|^2).
\end{equation}

\section{Environment Sensing Strategy  \label{vision-aided}}
In this section, we develop an environment sensing framework for dynamic scatterer detection and static environment reconstruction. Here, ``dynamic scatterers" broadly refer to moving or rotating objects, while ``static environment" encompasses long-term fixed objects and buildings, etc. It needs to be mentioned that the proposed framework can be extended to any indoor environment.
\begin{figure*}[!t]
	\centering
	
	\includegraphics[width=6.5in]{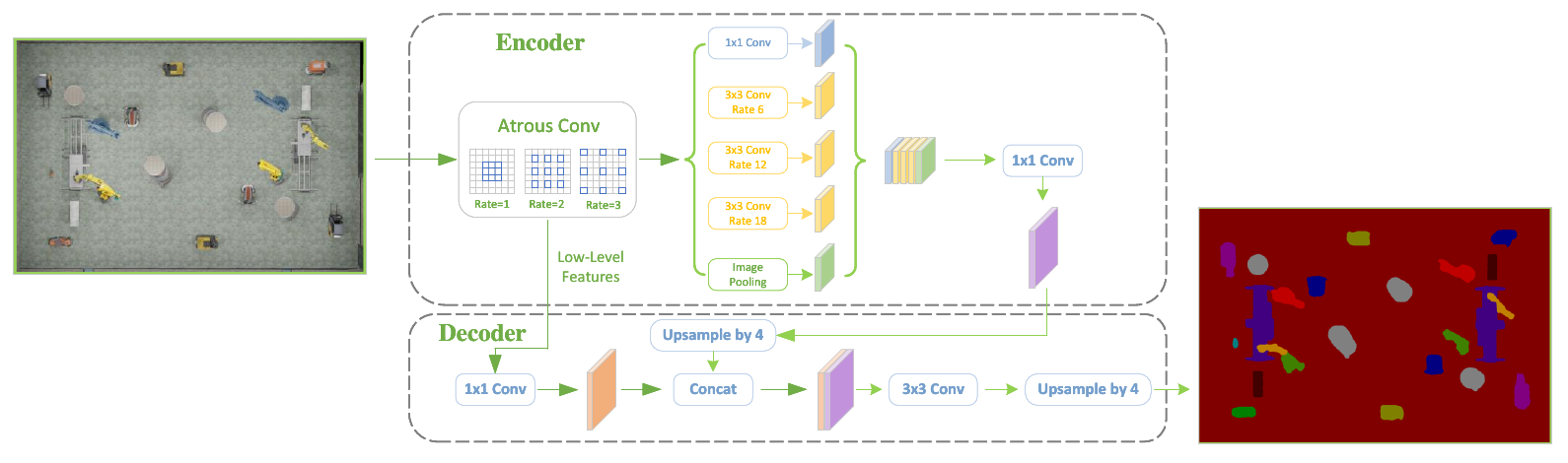}
	\caption{Diagram of the semantics segmentation using Deeplabv3+ model. The encoder module encodes multi-scale contextual information through atrous convolution at various scales, while decoder module refines the segmentation results along object boundaries.}
	\label{fig2}
\end{figure*}
\subsection{Dynamic Scatterers Detection}
We utilize red-green-blue (RGB) cameras to capture environmental images for dynamic scatterers detection. 
Considering that the environmental images contain both dynamic scatterers and static environment, we first utilize semantic
segmentation techniques~\cite{lateef2019survey} to obtain the segmentation maps of all objects within the environmental images. We utilize continuous frames segmentation maps to discern the motion status of each object, which could facilitate  the automatic removal of static environment while preserving dynamic scatterers.
Next, we design a 2D CNN network to extract dynamic features from dynamic scatterer segmentation maps.
\subsubsection{The semantic segmentation network\label{dy1}}
The semantic segmentation network Deeplabv3+~\cite{chen2018encoder} utilizes spatial pyramid pooling module and encode-decoder structure to achieve higher mean intersection over union (mIOU). Specifically, spatial pyramid pooling module is able to encode multi-scale contextual information by probing the incoming features with filters or pooling operations at multiple rates and multiple effective fields-of-view, while encode-decoder structure can capture sharper object boundaries by gradually recovering the spatial information. The Deeplabv3+ model combines the advantages of both structures to achieve high-precision semantic segmentation and sharper object boundaries.

Here, we utilize Deeplabv3+ model to segment environmental images, as shown in Fig.~\ref{fig2}.
Denote the environmental image captured by the $c$-th camera as $\boldsymbol{X}_c\in \mathbb{R}^{C \times H \times W}$, where $C,\  H$, and $W$ are the channel number, the height, and the width of RGB image respectively. Assuming there are $m$ rows and $n$ columns of cameras in the factory, a global environmental image  $\boldsymbol{X}\in \mathbb{R}^{C \times mH \times nW}$ can be obtained by stitching the images captured by all cameras. The label and the output of Deeplabv3+ model are denoted as $\boldsymbol{G}\in \mathbb{R}^{mH \times nW}$ and $\hat{\boldsymbol{ G}}\in \mathbb{R}^{Q \times mH \times nW}$, where the $(i,j)$-th entry of $\boldsymbol{G}$ is the label of the semantic class at the pixel coordinate $(i,j)$, i.e., $\boldsymbol{G}(i,j)=0, \cdots, Q-1$. Moreover, $Q$ is the number of semantic classes and $\hat{\boldsymbol{G}}(q, i, j)$ represents the probability that the $(i, j)$-th pixel belongs to the $q$-th class. The loss function of the Deeplabv3+ is cross entropy, which can be expressed as
\begin{equation}\mathrm{Loss}_{\{\mathrm{Dpv3+}\}}=-\frac{1}{mnH W} \sum_{i, j} \log \left(\frac{\exp (\hat{\boldsymbol{G}}(\boldsymbol{G}(i, j), i, j))}{\sum_{q=0}^{Q-1} \exp (\hat{\boldsymbol{G}}(q, i, j))}\right) .
\end{equation}
Using the stochastic gradient descent (SGD) algorithm~\cite{amari1993backpropagation} to minimize $\mathrm{Loss}_{\{\mathrm{Dpv3+}\}}$ until convergence, the segmentation map can be obtained from
\begin{equation}
	\boldsymbol{G}_{\mathrm{seg}}=\underset{q=0, \cdots, Q-1}{\arg \max } \hat{\boldsymbol{G}}(q, i, j).
\end{equation}
The performance of the semantics segmentation can be evaluated by the segmentation precision and the mIOU. The segmentation precision is given by
\begin{equation}
	A_{\mathrm{Dpv3+}}=\frac{\sum_{i, j} \sum_{n=1}^{N} \mathds{1}\left(\boldsymbol{G}_{\mathrm{seg}}^{(n)}(i, j)=\boldsymbol{G}^{(n)}{(i, j)}\right)}{mnH W N},
\end{equation}
where the superscript ${(n)}$ denotes the $n$-th sample, $N$ is the number of testing samples, and $\mathds{1}$ is the indicator function. The mIOU is given by
\begin{equation}
	\mathrm{mIOU} =\frac{1}{QN} \sum_{n=1}^{N}\sum_{q=0}^{Q-1} \frac{H^{(n)}_{q q}}{\sum_{p=0}^{Q-1} H_{q p}^{(n)}+\sum_{p=0}^{Q-1} H_{p q}^{(n)}-H_{q q}^{(n)}},
\end{equation}
where $H_{p q}^{(n)}$ represents the number of pixels that predict the $p$-th class as the $q$-th class. Based on $\boldsymbol{G}_{\mathrm{seg}}$, the segmentation maps of all classes such as ``engineering vehicle", ``mechanical arm", ``controller", ``machine tool", etc, can be easily separated by multiplying zero-masks.
\begin{algorithm}[!t]
	\caption{Two-pass algorithm for connected component analysis}
	\label{alg:1}
	\begin{algorithmic}[1]
		\REQUIRE $\boldsymbol{G}_{\mathrm{seg},q}\in\mathbb{R}^{mH\times nW}$
		\ENSURE  $\boldsymbol{G}_{\mathrm{seg},q,d}\in\mathbb{R}^{mH\times nW} , d=1,2,\cdots,D$
		\STATE \textbf{Step1: the first pass}
		\FOR{$i=1$ to $mH$}
		\FOR{$j=1$ to $nW$}
		\IF{$\boldsymbol{G}_{\mathrm{seg},q}(i,j)=q$} 
		\IF{$\boldsymbol{G}_{\mathrm{seg},q}(i-1,j)=0$ and $\boldsymbol{G}_{\mathrm{seg},q}(i,j-1)=0$}
		\STATE $\mathrm{label} = \mathrm{label} + 1$, $\boldsymbol{G}_{\mathrm{seg},q}(i,j)=\mathrm{label}$
		\ELSIF{$\boldsymbol{G}_{\mathrm{seg},q}(i-1,j)>1$ or $\boldsymbol{G}_{\mathrm{seg},q}(i,j-1)>1$}
		\STATE $\boldsymbol{G}_{\mathrm{seg},q}(i,j)=\min\{\boldsymbol{G}_{\mathrm{seg},q}(i-1,j),$  $\boldsymbol{G}_{\mathrm{seg},q}(i,j-1)\}$
		\STATE Record the connectivity area set $\mathrm{labelSet}[k]$ to which $\mathrm{label}$ belongs
		\ENDIF
		\ENDIF
		\ENDFOR
		\ENDFOR
		\STATE \textbf{Step2: the second pass}
		\FOR{$i=1$ to $mH$}
		\FOR{$j=1$ to $nW$}
		\IF{$\boldsymbol{G}_{\mathrm{seg},q}(i,j)>1$ and $\boldsymbol{G}_{\mathrm{seg},q}(i,j)\in \mathrm{labelSet}[k]$}
		\STATE $\boldsymbol{G}_{\mathrm{seg},q}(i,j)=k$
		\ENDIF
		\ENDFOR
		\ENDFOR
		\STATE $\boldsymbol{G}_{\mathrm{seg},q,d} = \mathds{1}( \boldsymbol{G}_{\mathrm{seg},q}=d) 
		\odot \boldsymbol{G}_{\mathrm{seg},q}$
	\end{algorithmic}
\end{algorithm}

\subsubsection{Instance separation and refinement\label{dy2}}
Denote the $q$-th class segmentation map as $\boldsymbol{G}_{\mathrm{seg},q}$, which can be expressed as 
\begin{equation}
	\boldsymbol{G}_{\mathrm{seg},q} = \boldsymbol{M}_{q} \odot \boldsymbol{G}_{\mathrm{seg}},
\end{equation}
where $\boldsymbol{M}_{q} = \mathds{1}( \boldsymbol{G}_{\mathrm{seg}}=q), q=0, \cdots, Q-1$ is the zero-mask of the $q$-th class, and $\odot$ represents the Hadamard product of a matrix. Since the same class of semantics often contains multiple instances and each instance is usually a connected entity, we propose to utilize the two-pass algorithm~\cite{wu2009optimizing} to separate and refine instances within the same class. The core idea of two-pass algorithm is that the first pass assigns different label values for the pixels of interest and records the label sets to which the label belongs, while the second pass unifies the different labels in the same label set to separate each connected instance. The detailed steps of two-pass algorithm are given in Algorithm~\ref{alg:1}. 
\begin{algorithm}[!t]
	\caption{Instance refinement}
	\label{alg:2}
	\begin{algorithmic}[1]
		\REQUIRE  $\boldsymbol{G}_{\mathrm{seg},q,d}\in\mathbb{R}^{mH\times nW} , d=1,2,\cdots,D$
		\ENSURE   $\boldsymbol{G}_{\mathrm{seg},q,d}\in\mathbb{R}^{mH\times nW} , d=1,2,\cdots,D^{'}$
		\STATE Initialize pixel ratio $\mu$
		\STATE Count the pixel number of $D$ instances: $\boldsymbol{n}_p = [N_{p,1},N_{p,2},\cdots,N_{p,D}]$
		\WHILE{True}
		\STATE $N_{p,min}=\min\{\boldsymbol{n}_p\}$, $I_{min} = \arg\min \{\boldsymbol{n}_p\}$
		\IF{$\frac{N_{p,min}}{(\sum_{i=1}^{\rm{dim}(\boldsymbol{n}_p)}N_{p,i}-N_{p,min})/(\rm{dim}(\boldsymbol{n}_p)-1)} < \mu$} 
		\STATE $D = D - 1$, remove $\boldsymbol{G}_{\mathrm{seg},q,I_{min}}$ and $\boldsymbol{n}_p(I_{min})$
		\ELSE
		\STATE break
		\ENDIF
		\ENDWHILE
		\STATE $D^{'} = D$, $\boldsymbol{G}_{\mathrm{seg},q,d}, d=1,2,\cdots,D^{'}$
	\end{algorithmic}
\end{algorithm}
\begin{figure}[!t]
	\centering
	
	\includegraphics[width=3.5in]{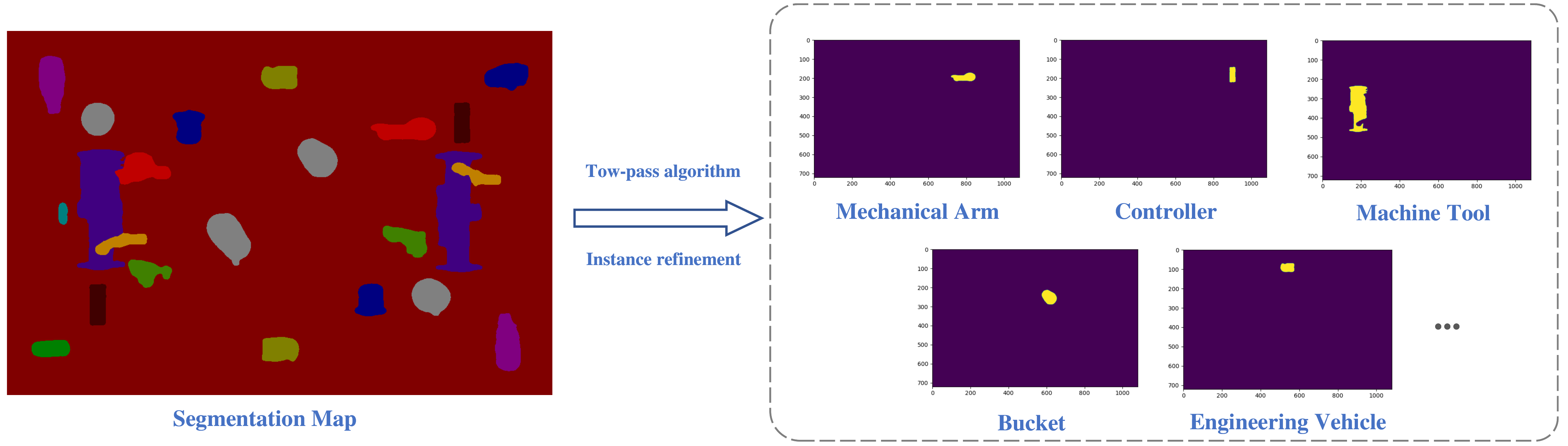}
	\caption{Illustration of the instance separation and refinement.}
	\label{fig3}
\end{figure}

Assuming that there are $D$ connected entities in the $q$-th semantic class maps, we can obtain $D$ segmentation maps $\boldsymbol{G}_{\mathrm{seg},q,d}\in\mathbb{R}^{mH\times nW} , d=1,2,\cdots,D$. Due to the errors of semantic segmentation model, the $q$-th class of semantics actually only contains $D^{'}$ instances $(D^{'}<D)$, and thus we need to refine the results of the two-pass algorithm. 
Considering that semantic segmentation errors often yield a small range of connected entities, we count the pixel  number of instance in $D$ segmentation maps and select the map with the smallest pixel number of instance each time. If the pixel number is significantly fewer than the mean of other instance pixel number, then the instance will be considered as an error and will be deleted. 
The detailed algorithm to determine whether the instance is generated by semantic segmentation errors is shown in Algorithm~\ref{alg:2}.

\subsubsection{Dynamic scatterers separation\label{dy3}}
\begin{figure*}[!t]
	\centering
	\includegraphics[width=6.5in]{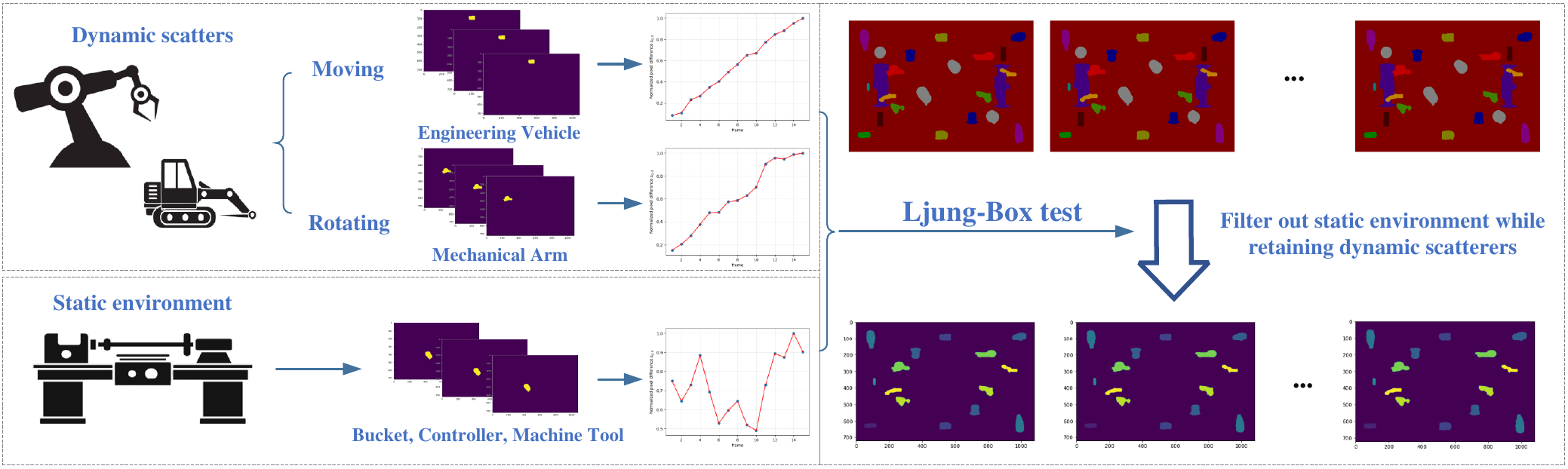}
	\caption{Illustration of the dynamic scatterer separation. Based on the autocorrelation of difference sequence between the dynamic scatterers and the static environment, the dynamic scatterers will be separated from the static environment.}
	\label{fig4}
\end{figure*}

The set of segmentation maps for the $d$-th instance of the $q$-th class with consecutive $l$ frames can be represented by
\begin{equation}
	\mathcal{G}_{q,d} = \{\boldsymbol{G}_{\mathrm{seg},q,d}^{(n)}, \boldsymbol{G}_{\mathrm{seg},q,d}^{(n+1)}, \cdots, \boldsymbol{G}_{\mathrm{seg},q,d}^{(n+l-1)}\},
\end{equation}
where $d=1,2,\cdots,D^{'}, q=1,2,\cdots,Q$. Considering that the locations or angles of dynamic scatterers vary regularly over time while static scatterers remain stationary, we propose to utilize the difference of continuous frame segmentation maps to separate dynamic scatterers. The number of different pixel values between the $(n+t)$-th frame and the $n$-th frame for the $d$-th instance map of the $q$-th class can be expressed as
\begin{equation}
	s_{q,d}^{(n+t)} = \sum_{i=1}^{mH}\sum_{j=1}^{nW}\mathds{1}(\|\boldsymbol{G}_{\mathrm{seg},q,d}^{(n+t)}(i,j)- \boldsymbol{G}_{\mathrm{seg},q,d}^{(n)}(i,j)\|>0),
\end{equation}
where $t=1,2,\cdots,l-1$. Then we can obtain a difference sequence $\{s_{q,d}^{(n+t)}\}_{t=1,2,\cdots,l-1} $, where the length of $\{s_{q,d}^{(n+t)}\}_{t=1,2,\cdots,l-1} $ is $l-1$.  Due to the continuity of the rotation or movement of dynamic scatterers, the difference sequence of dynamic scatterers will exhibit strong autocorrelation, while the difference sequence of static environment due to semantic segmentation errors can be considered as white noise with zero autocorrelation coefficient, as shown in Fig~\ref{fig4}. Therefore, we propose to utilize the Ljung-Box test method~\cite{ljung1978measure} to statistically separate the dynamic scatterers and static environment. The Ljung-Box test uses the autocorrelation coefficients of samples to construct statistics that follow a specific distribution, and then uses hypothesis testing to determine whether a series of observations within a certain time period are white noise. If the number of different pixel values between the $(n+t)$-th frame and the $n$-th frame  for the $d$-th instance map of the $q$-th class is a random variable $S_{q,d}^{(n+t)}$, then continuous $l-1$ random variables can form a discrete stochastic process $\{S_{q,d}^{(n+t)}\}_{t=1,2,\cdots,l-1}$. The sample $\{s_{q,d}^{(n+t)}\}_{t=1,2,\cdots,l-1} $ is a realization of stochastic process $\{S_{q,d}^{(n+t)}\}_{t=1,2,\cdots,l-1} $. Assuming $\{S_{q,d}^{(n+t)}\}_{t=1,2,\cdots,l-1} $ is a wide-sense stationary stochastic process~\cite{stationary}, the autocorrelation coefficient with an interval of $r$ can be expressed as
\begin{equation}
	\begin{aligned}
		\rho_r&=\frac{\operatorname{Cov}\left(S_{q,d}^{(n+t)}, S_{q,d}^{(n+t-r)}\right)}{\sqrt{\operatorname{Var}\left(S_{q,d}^{(n+t)}\right) \operatorname{Var}\left(S_{q,d}^{(n+t-r)}\right)}}\\
		&=\frac{\operatorname{Cov}\left(S_{q,d}^{(n+t)},S_{q,d}^{(n+t-r)}\right)}{\operatorname{Var}\left(S_{q,d}^{(n+t)}\right)},
	\end{aligned}
\end{equation}
where $\operatorname{Cov}$ and $\operatorname{Var}$ represent the covariance and variance of random variables. However, the probability distribution that $S_{q,d}^{(n+t)}$ follows is often difficult to obtain, and in practice, only one sample $\{s_{q,d}^{(n+t)}\}_{t=1,2,\cdots,l-1} $ can be obtained. Then, the autocorrelation coefficient of the sample with an interval of $r$ can be expressed as
\begin{algorithm}[!t]
	\caption{Dynamic scatterers detection algorithm}
	\label{alg:3}
	\begin{algorithmic}[1]
		\REQUIRE Environment image sequence $\boldsymbol{X}^{(n)},\boldsymbol{X}^{(n+1)}, \cdots,$ $\boldsymbol{X}^{(n+l-1)}.$ 
		\ENSURE  Dynamic scatterers map sequence  $\boldsymbol{D}^{(n)},$
		$\boldsymbol{D}^{(n+1)}, \cdots, \boldsymbol{D}^{(n+l-1)}.$ 
		\STATE \textbf{Step1: Semantic segmentation}
		\STATE  $\boldsymbol{G}_{\mathrm{seg}}^{(n)},\boldsymbol{G}_{\mathrm{seg}}^{(n+1)}, \cdots, \boldsymbol{G}_{\mathrm{seg}}^{(n+l-1)}$ $\leftarrow$ $\boldsymbol{X}^{(n)},\boldsymbol{X}^{(n+1)}, \cdots,$ $\boldsymbol{X}^{(n+l-1)}.$ 
		\STATE \textbf{Step2: Instance separation and refinement}
		\FOR{$t=0$ to $l-1$}
		\FOR{$q=1$ to $Q$}
		\STATE $\boldsymbol{G}^{(n+t)}_{\mathrm{seg},q} = \boldsymbol{M}^{(n+t)}_{q} \odot \boldsymbol{G}^{(n+t)}_{\mathrm{seg}}$
		\STATE Two-pass algorithm: $\boldsymbol{G}^{(n+t)}_{\mathrm{seg},q,d}\leftarrow \boldsymbol{G}^{(n+t)}_{\mathrm{seg},q} , d=1,2,\cdots,D$
		\STATE Instance refinement: $ \boldsymbol{G}^{(n+t)}_{\mathrm{seg},q,d}(d=1,2,\cdots,D^{'})\leftarrow\boldsymbol{G}^{(n+t)}_{\mathrm{seg},q,d}(d=1,2,\cdots,D) $
		\ENDFOR
		\ENDFOR
		\STATE \textbf{Step3: Dynamic semantics separation}
		\FOR{$q=1$ to $Q$}
		\FOR{$d=1$ to $D^{'}$}
		\FOR{$t=1$ to $l-1$}
		\STATE $s_{q,d}^{(n+t)} = \sum_{i=1}^{mH}\sum_{j=1}^{nW}\mathds{1}(\boldsymbol{G}_{\mathrm{seg},q,d}^{(n+t)}(i,j)- \boldsymbol{G}_{\mathrm{seg},q,d}^{(n)}(i,j)>0)$
		\ENDFOR
		\STATE $\hat{\rho}_r=\frac{\sum_{t=r+1}^{T}\left(s_{q,d}^{(n+t)}-\bar{s}_{q,d}\right)\left(s_{q,d}^{(n+t-r)}-\bar{s}_{q,d}\right)}{\sum_{t=1}^{T}\left(s_{q,d}^{(n+t)}-\bar{s}_{q,d}\right)^2},$
		$ r=1,2,\cdots,m. $
		\STATE $Q(m)=T(T+2) \sum_{j=1}^m \frac{\hat{\rho}_j^2}{T-j}$
		\IF{ $Q(m)\leq \chi^2_{1-\alpha}(m)$}
		\STATE Remove $\boldsymbol{G}_{\mathrm{seg},q,d}^{(n)},\boldsymbol{G}_{\mathrm{seg},q,d}^{(n+1)}, \cdots, \boldsymbol{G}_{\mathrm{seg},q,d}^{(n+l-1)}.$
		\ENDIF
		\ENDFOR
		\ENDFOR
		\STATE  $\boldsymbol{D}^{(n)},\boldsymbol{D}^{(n+1)}, \cdots, \boldsymbol{D}^{(n+l-1)}\leftarrow \boldsymbol{G}_{\mathrm{seg}}^{(n)},\boldsymbol{G}_{\mathrm{seg}}^{(n+1)}, \cdots,$
		$ \boldsymbol{G}_{\mathrm{seg}}^{(n+l-1)}$ 
	\end{algorithmic}
\end{algorithm}
\begin{equation}
	\begin{aligned}
	\hat{\rho}_r=\frac{\sum_{t=r+1}^{T}\left(s_{q,d}^{(n+t)}-\bar{s}_{q,d}\right)\left(s_{q,d}^{(n+t-r)}-\bar{s}_{q,d}\right)}{\sum_{t=1}^{T}\left(s_{q,d}^{(n+t)}-\bar{s}_{q,d}\right)^2}, \\ \quad 0 < r<T-1,
	\end{aligned}
\end{equation}
where $T=l-1$ is the length of the sequence and $\bar{s}_{q,d}=\sum_{t=1}^{T} s_{q,d}^{(n+t)} / T$ is the mean of the samples. If the elements of sequence $\{s_{q,d}^{(n+t)}\}_{t=1,2,\cdots,T} $ are independent of each other, then there are $\hat{\rho}_1=\hat{\rho}_2= \cdots =\hat{\rho}_m=0$; otherwise  $\hat{\rho}_1, \hat{\rho}_2, \cdots, \hat{\rho}_m$ are not all zero. Therefore, we can formulate  the following null hypothesis and alternative hypothesis:
\begin{equation}\begin{array}{l}
		H_0: \hat{\rho}_1=\hat{\rho}_2= \cdots =\hat{\rho}_m=0;\\
		H_1: \hat{\rho}_1, \hat{\rho}_2, \cdots, \hat{\rho}_m \quad \text{are not all zero.}
\end{array}\end{equation} 
The statistics $Q(m)$ proposed by Ljung and Box~\cite{ljung1978measure} asymptotically follow the $m$-order chi square distribution~\cite{lancaster2005chi} $\chi^2(m)$ and $Q(m)$ can be expressed as
\begin{equation}
	Q(m)=T(T+2) \sum_{j=1}^m \frac{\hat{\rho}_j^2}{T-j}.
\end{equation}
The decision rule is that when $Q(m)>\chi^2_{1-\alpha}(m)$, $H_0$ will be rejected. The sequence is considered to exhibit autocorrelation and the instance is a dynamic scatterer. When $Q(m)\leq \chi^2_{1-\alpha}(m)$, $H_0$ will be accepted. The sequence is considered as white noise and the instance belongs to static environment. The entire process of dynamic scatterer separation is shown in Algorithm \ref{alg:3}.

\subsubsection{Dynamic feature extraction} The residual structures facilitate the training of deeper networks by mitigating the vanishing gradient and enabling effective feature reuse across layers~\cite{he2016deep}. Here, we take the dynamic scatterer segmentation maps extracted from Section \ref{dy3} as input and design a network with residual structure stacked by 2D convolution, BatchNorm, and ReLU to extract dynamic features. After the feature layer is flattened, a fully connected network is added to the output layer to obtain a dynamic feature vector. The detailed dynamic feature extraction network structure is shown in Tab. \ref{table1}. Denote the mathematical function of the dynamic feature extraction module as $f_{\{\mathrm{dy}\}}$. Then the dynamic feature of the $n$-th frame can be expressed as
\begin{equation}
	\boldsymbol{d}^{(n)} = f_{\{\mathrm{dy}\}}(\boldsymbol{D}^{(n)}),
\end{equation}
where $\boldsymbol{d}^{(n)} \in \mathbb{R}^{N_D \times 1}$ and $N_D$ is the number of neurons in the output layer of module $ f_{\{\mathrm{dy}\}}$.
\begin{table*}[!h]
	\caption{The network parameters of dynamic, static, user identification feature extraction\label{table1}}
	\centering
	\tabcolsep=0.3cm
	\begin{tabular}{|c|c|c|c|c|c|}
		\hline
		Module & Layer & Kernel & Stride & Neuron & Output Shape \\
		\hline
		\multirow{5}{*}{Dynamic-feature} & [Covn-BatchNorm-ReLU]  &7 & 2 & 32 & 32$\times$112$\times$168 \\
		& MaxPool & 3 &2 & / & 32$\times$56$\times$84\\
		& [Covn-BatchNorm-ReLU]$\times$2 $<$Residual$>$  &3,3 & 1,1 & 64, 64& 64$\times$56$\times$84 \\
		& [Covn-BatchNorm-ReLU]$\times$2 $<$Residual$>$  &3,3 & 2,1 & 128, 128 & 128$\times$28$\times$42 \\
		& [AdaptiveAvgPool2d-FC] &/ & / & 128 & 128 \\
		\hline
		\multirow{5}{*}{Static-feature} & [Covn(3D)-BatchNorm-ReLU]  &3 & 1 & 32 & 32$\times$31$\times$21$\times$11 \\
		& AvgPool & 3 &2 & / & 16$\times$20$\times$40\\
		& [Covn(3D)-BatchNorm-ReLU]$\times$2 $<$Residual$>$  &3,3 & 1,1 & 64, 64 & 64$\times$31$\times$21$\times$11 \\
		& [Covn(3D)-BatchNorm-ReLU]$\times$2 $<$Residual$>$  &3,3 & 2,1 & 128, 128 & 128$\times$16$\times$11$\times$6 \\
		& [AdaptiveAvgPool3d-FC] &/ & / & 116 & 116\\
		\hline
		\multirow{2}{*}{User-identification-feature} & BatchNorm  &/ & / &3 & 3 \\
		& [FC-BatchNorm-Tanh] $\times$ 2 & / &/ & 32, 12 &  12 \\
		\hline
	\end{tabular}
\end{table*}

\subsection{Static Environment Reconstruction}
Static environment mainly includes steel crossbeams, steel columns, machine tools, controllers, factory building, etc. The downward-looking cameras can only capture objects on the floor, while static environment such as steel crossbeams at the ceiling of the factory and the structure of buildings cannot be captured by cameras. Therefore, we utilize a LIDAR equipped on the ceiling to capture 3D point clouds of the static environment and convert the point cloud into fixed dimensional pseudo image sequences by a grid based approach. Next, we design a 3D CNN network to extract static features from pseudo image sequences.
\subsubsection{Point cloud to Pseudo Image sequence}
The scanned point cloud by LIDAR can be represented as a set $\mathcal{P}=\{ \boldsymbol{p}_1,\boldsymbol{p}_2,\cdots,\boldsymbol{p}_G \}$, where $G$ denotes the number of points in the point cloud and each element $\boldsymbol{p}_g\in \mathbb{R}^{3\times1}$ represents the three-dimensional coordinates of one point.
Due to the varying number of $G$ generated each time, we cannot use CNN structure for feature extraction. Therefore, we propose to utilize a grid based approach to convert point clouds into fixed dimensional pseudo image sequences. Specifically, we can find a suitable cube $\boldsymbol{\mathcal{C}}$ that includes the entire static environment. Then the length, the width, and the height of the cube $\boldsymbol{\mathcal{C}}$ can be evenly divided into $a, b,$ and $c$ parts respectively to obtain $a\times b\times c$ small cubes. We establish a local coordinate system with the center point of each small cube as the origin, and convert all points contained in this small cube to the local coordinate system. 
Denote the $(i,j,k)$-th small cube as $\boldsymbol{\mathcal{C}}_{i,j,k}$, and then the point cloud set contained in $\boldsymbol{\mathcal{C}}_{i,j,k}$ after coordinate system transformation can be expressed as
\begin{equation}
 	\mathcal{P}_{i,j,k}=  \{ \boldsymbol{p}_g - \bar{\boldsymbol{c}}_{i,j,k} |\boldsymbol{p}_g \in  \boldsymbol{\mathcal{C}}_{i,j,k} \}
\end{equation}
where  $\bar{\boldsymbol{c}}_{i,j,k}$ is the three-dimensional coordinate of the center point of $\boldsymbol{\mathcal{C}}_{i,j,k}$. In order to convert the point cloud into a fixed dimensional format without changing its spatial structure, we use the mean value of local coordinates of all points within $\boldsymbol{\mathcal{C}}_{i,j,k}$ as a representation of these points, i.e., $\bar{\boldsymbol{p}}_{i,j,k}= \frac{1}{|\mathcal{P}_{i,j,k}|}  \sum_g \boldsymbol{p}_g, \boldsymbol{p}_g\in\mathcal{P}_{i,j,k}.$ If there is no point in $\boldsymbol{\mathcal{C}}_{i,j,k}$, then $\bar{\boldsymbol{p}}_{i,j,k}$ is set to 0. We partition the
$G$ points into $a\times b\times c$ spatial blocks and obtain the pseudo image sequence $\boldsymbol{P} \in \mathbb{R}^{a\times b\times c\times 3}$, where $a,b, c,$ and ``3" are equivalent to the length of sequence, the height, the width, and the number of channels of a pseudo image, respectively. Compared with the original point cloud $\mathcal{P}$, the dimension of pseudo image sequence $\boldsymbol{P}$ is significantly reduced, and the fixed dimensional data format makes it convenient to use 3D CNNs for feature extraction. Note that the grid size can be adjusted according to the complexity of the actual static environment. When the environment is complex, we could reduce the grid size to obtain more detailed features, and when the environment is simple, we could increase the grid size to reduce computational complexity.

\subsubsection{Static feature extraction}
The traditional 2D CNN can only extract features from a single image, but cannot extract temporal correlation information from consecutive frame image sequences. In comparison, the 3D CNN can extract relevant information between consecutive frames by introducing time dimension, and has wide applications in video classification and  behavior recognition~\cite{maturana2015voxnet}. Hence, we utilize 3D CNN to extract spatial features of pseudo image sequence $\boldsymbol{P} \in \mathbb{R}^{a\times b\times c\times 3}$. Similar to dynamic feature extraction, we take $\boldsymbol{P}$ as input and design a residual structured network stacked by 3D convolution, BatchNorm, and ReLU to extract static features. After the feature map sequence is flattened, a fully connected network is added to the output layer to obtain a static feature vector. The detailed static feature extraction network structure is shown in Tab. \ref{table1}. Denote the mathematical function of the static feature extraction module as $f_{\{\mathrm{st}\}}$. Then the static feature  can be expressed as
\begin{equation}
	\boldsymbol{s} = f_{\{\mathrm{st}\}}(\boldsymbol{P}),
\end{equation}
where $\boldsymbol{s} \in \mathbb{R}^{N_S \times 1}$ and $N_S$ is the number of neurons in the output layer of module $ f_{\{\mathrm{st}\}}$.

\subsection{User Identification Feature Extraction} Dynamic and static features contain all the information of the communication environment, but the UE at different locations will correspond to different channels. Therefore, user identification information needs to be used as the input to distinguish user at different locations. Considering that each equipment in the smart factory is under control, we assume that the user's location is precisely known at BS and will be used as user identification information\footnote{In other indoor environments, user identification information can be distinguished using historical beam indices or other predetermined information.}.  Denoting the location of the $n$-th frame of the user as $\boldsymbol{l}^{(n)} \in \mathbb{R}^{3\times 1}$, the identification module is composed of several fully connected (FC) blocks, and each FC block is sequentially stacked by a FC, a BarchNorm, and a Tanh layers. The detailed network structure is shown in Tab.~\ref{table1}. Denote the mathematical function of the user identification feature extraction module as $f_{\{U\}}$. Then the user identification feature of the $n$-th frame can be expressed as
\begin{equation}
	\boldsymbol{u}^{(n)} = f_{\{\mathrm{id}\}}(\boldsymbol{l}^{(n)}),
\end{equation}
where $\boldsymbol{u}^{(n)} \in \mathbb{R}^{N_U \times 1}$ and $N_U$ is the number of neurons in the output layer of module $ f_{\{\mathrm{id}\}}$.
\begin{figure*}[!t]
	\centering
	\includegraphics[width=6.5in]{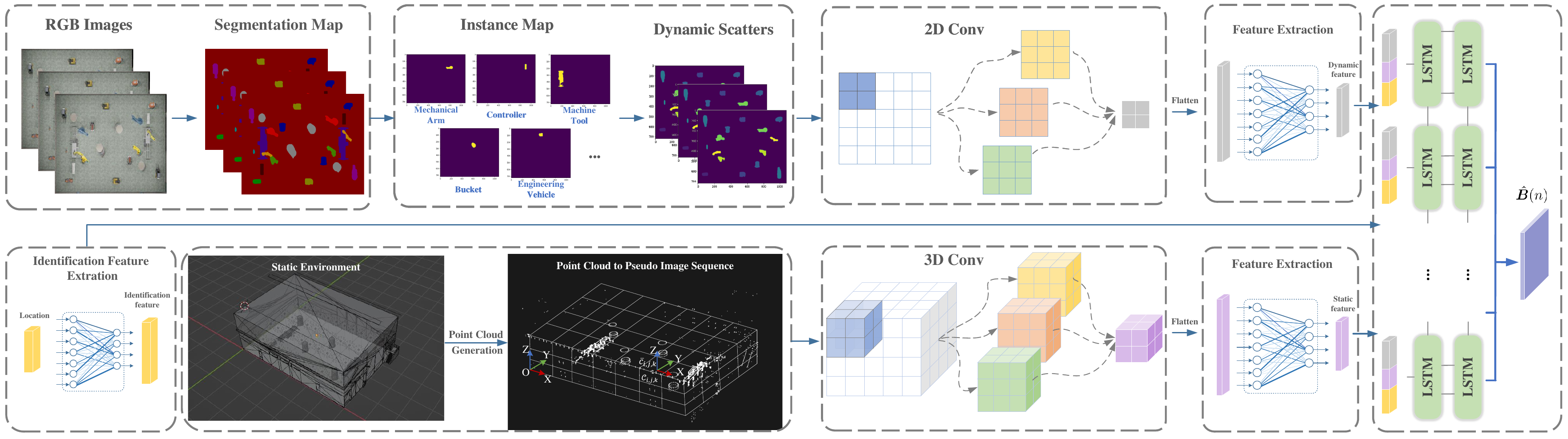}
	\caption{Diagram of the environment sensing-aided beam prediction network architecture. The dynamic feature, the static feature, and the user identification feature are extracted and then fed into LSTM networks to predict the optimal beam indices in advance.}
	\label{fig6}
\end{figure*}
\section{Beam Prediction and Transfer Learning Strategy \label{trans}}
In this section, we propose a pre-training model for beam prediction and design a transfer learning strategy to generalize the proposed pre-trained model from given environments to a new environment.
\subsection{Beam Prediction}
We propose to use continuous $l$ frames of environmental images, UE's locations, and static environmental point clouds to predict the beam pair indices of the $k$-th future frame  between $B$ BSs and the UE. Denote the dynamic feature and the user identity feature of the $n$-th frame in the $e$-th environment as $\boldsymbol{d}^{(n)}_e\in \mathbb{R}^{N_D\times1}$ and $\boldsymbol{u}^{(n)}_e\in \mathbb{R}^{N_U\times1}$,  respectively and denote the static feature of the $e$-th environment as $\boldsymbol{s}_e\in \mathbb{R}^{N_S\times1}$, where $e\in\{1,2,\cdots,E\}$. Next we fuse these three features to form sensing feature $\boldsymbol{f}_e^{(n)} = [(\boldsymbol{u}^{(n)}_e)^T, (\boldsymbol{d}^{(n)}_e)^T, \boldsymbol{s}_e^T]^T\in \mathbb{R}^{(N_U+ N_D + N_S)\times 1}$. Due to the strong temporal relationship in the prediction process, we utilize LSTM networks to predict the beam pair indices for the $k$-th future  frame. By introducing forget gate, input gate, and output gate, LSTM can greatly improve the ability of recurrent neural network (RNN) to process time-series information with long time dependence~\cite{hochreiter1997long}. The sensing features of consecutive $l$ frames in the $e$-th environment will be fed into a two-layer LSTM network with $l$ LSTM cells. Considering that beam prediction involves selecting the optimal beam pair from $N_tN_r$ beam pairs, which is essentially a multi-classification problem, we add a softmax function to the output layer. To ease the representation, the dynamic scatterers detection and static environment reconstruction are recorded as operator $\rm DS(\cdot)$ and $\rm SR(\cdot)$, and the two-layer LSTM with $l$ cells is recorded as operator ${\rm LMST}_{(l)}^{(2)}(\cdot)$. The overall beam prediction algorithm can be expressed as
\begin{subequations}
	\begin{align}
			&\{\boldsymbol{D}_e^{(n)},\boldsymbol{D}_e^{(n+1)}, \cdots, \boldsymbol{D}_e^{(n+l-1)}\}\nonumber\\
			&=\mathrm{DS}(\{\boldsymbol{X}_e^{(n)},\boldsymbol{X}_e^{(n+1)},\cdots, \boldsymbol{X}_e^{(n+l-1)}\}),\\
			&\{\boldsymbol{d}_s^{(n)}, \boldsymbol{d}_s^{(n+1)}, \cdots, \boldsymbol{d}_s^{(n+l-1)}\} \nonumber\\
			&= f_{\{\mathrm {dy}\}}(\{\boldsymbol{D}_e^{(n)},\boldsymbol{D}_e^{(n+1)}, \cdots, \boldsymbol{D}_e^{(n+l-1)}\}),\\
			&\boldsymbol{P}_e = \mathrm{SR}(\mathcal{P}_e) ,\\
			&\boldsymbol{s}_e = f_{\{\mathrm{st}\}}(\boldsymbol{P}_e),
	\end{align}
\end{subequations}
\begin{align}
			&\{\boldsymbol{u}_s^{(n)}, \boldsymbol{u}_s^{(n+1)}, \cdots, \boldsymbol{u}_s^{(n+l-1)}\} \nonumber\\
			&= f_{\{\mathrm{id}\}}(\{\boldsymbol{l}_e^{(n)},\boldsymbol{l}_e^{(n+1)}, \cdots, \boldsymbol{l}_e^{(n+l-1)}\}),\tag{21e}\\
			&\boldsymbol{f}_e^{(n)} = [(\boldsymbol{u}^{(n)}_e)^T, (\boldsymbol{d}^{(n)}_e)^T, \boldsymbol{s}_e^T]^T, \cdots,\nonumber\\
			&\boldsymbol{f}_e^{(n+l-1)} = [(\boldsymbol{u}^{(n+l-1)}_e)^T, (\boldsymbol{d}^{(n+l-1)}_e)^T, \boldsymbol{s}_e^T]^T\tag{21f}\\
			&\hat{\boldsymbol{B}}(n)={\rm SoftMax}({\rm Reshape}({\rm FC}_{(N_tN_rB)}(\operatorname{LSTM}_{(l)}^{(2)}(\{\boldsymbol{f}_e^{(n)},\nonumber \\ 
			&\boldsymbol{f}_e^{(n+1)}, \cdots, \boldsymbol{f}_e^{(n+l-1)}\}))),\tag{21g}
\end{align}
where ${\rm FC}_{(N_tN_rB)}(\cdot)$ represents the FC layer with $N_tN_rB$ output neurons. The reshape function ${\rm Reshape}(\cdot)$ converts the $N_tN_rB\times 1$ dimensional vector output by the FC blocks into the $N_tN_r\times B$ dimensional matrix. Moreover, matrix-form softmax function can be expressed as 
\begin{multline}
	{\rm SoftMax}([\hat{\boldsymbol{b}}_1,\hat{\boldsymbol{b}}_2, ..., \hat{\boldsymbol{b}}_B]) = [   
		\frac{\exp(\hat{\boldsymbol{b}}_1)}{\sum_{j=1}^{N_tN_r}{\rm exp}(\hat{b}_1^j)},\\ \frac{\exp(\hat{\boldsymbol{b}}_2)}{\sum_{j=1}^{N_tN_r}{\rm exp}(\hat{b}_2^j)}, \cdots, \frac{\exp(\hat{\boldsymbol{b}}_B)}{\sum_{j=1}^{N_tN_r}{\rm exp}(\hat{b}_{B}^j)}
	],
\end{multline}
where $\hat{\boldsymbol{b}}_i$ is the $i$-th column of $\hat{\boldsymbol{B}}\in \mathbb{R}^{N_tN_r\times B}$ and $\hat{b}_i^j$ is the $j$-th element of $\hat{\boldsymbol{b}}_i$.
The softmax function coverts $\hat{\boldsymbol{b}}_i(n)$ into a probability vector, where $\hat{b}_i^j$ represents the probability that the optimal beam pair between the $i$-th BS and the UE is the $j$-th beam pair. Moreover, we utilize one-hot vectors  $\boldsymbol{b}_i(n)$ to form training labels  $\boldsymbol{ {B}}(n)\in \mathbb{R}^{N_tN_r\times B}$. If the optimal beam pair between the $i$-th BS and the UE is the $j$-th beam pair, then there are $b_i^j(n)=1$ and $b_i^k(n)=0,k\neq j$. The cross entropy commonly used by multi-classification tasks is selected as the loss function, i.e.,
\begin{equation}
	\label{corss}
	{\rm Loss}(\boldsymbol{ {B}}(n), \hat{\boldsymbol{ B}}(n)) = -\sum_{i=1}^{B}\sum_{j=1}^{N_tN_r}b_i^j(n) \log \hat b_i^j(n).
\end{equation}

\subsection{Transfer Learning Strategy}

The design of independent frameworks for dynamic and static feature extraction aims to adapt to variations of communication environments.  
For instance, when a factory produces a certain type of product, dynamic scatterers and users move within a fixed static environment. However, when the factory switches to producing a new type of product, dynamic scatterers and users navigate within a new static environment. In such cases, we can utilize LIDAR to rescan the new static environment to obtain 3D point clouds and capture environmental images during factory production to achieve rapid collection of environmental information.
In other indoor environments, such as conference rooms, the static environment includes buildings and furniture like sofas, tables, whereas dynamic scatterers are moving humans within the conference room. Therefore, the proposed framework is applicable to any indoor environment and aims to overcome the transfer challenge of sensing-aided wireless communication models in a new static environment with dynamic scatterers.

At the level of environmental sensing, dynamic features, static features, and user identity features already cover complete environmental and channel information. However, at the level of machine learning, a mapping model learned only from a limited number of environments in the training set often overfits these environments and lack the ability to generalize to a new environment. In real world, the variations of the environment are unpredictable, which means that the sample space of multiple environments is very large, and it is not possible to collect all potential environment data. Therefore, we propose a transfer learning strategy that can expeditiously adapt to new environments with limited labeled data.  

Denote the space\footnote{The space of dynamic scatterers $\mathcal{D}_e$ means the set of all possible dynamic scatterer  maps $\boldsymbol{D}_e^{(n)}$.} of dynamic scatterers  and the space\footnote{The space of user identification $\mathcal{U}_e$  means the set of all possible user's locations $\boldsymbol{l}_e^{(n)}$.} of user identification in the $e$-th environment as $\mathcal{D}_e$ and $\mathcal{U}_e$, respectively. The static environment can be represented by 3D point clouds $\mathcal{P}_e$. Define the space of the $e$-th environment feature and the space of beam indices as $\mathcal{X}(e) = \{\mathcal{D}_e, \mathcal{P}_e, \mathcal{U}_e\}$ and $\mathcal{Y}(e)$. Then, the problem of transfer learning can 
be defined as follows: i) Given $E_s$ source environment features $\{\mathcal{X}_{S}(e)\}_{e=1}^{E_s}$ and $E_s$ source tasks $\{Y(e)^{S}\}_{e=1}^{E_s}$, along with the target environment feature $\mathcal{X}_{T}$, and the target task $\mathcal{Y}_{T}$. ii) The objective is to leverage the knowledge gained from $\{\mathcal{X}_{S}(e)\}_{e=1}^{E_s}$ and $\{Y(e)^{S}\}_{e=1}^{E_s}$ to enhance the performance of the target task $\mathcal{Y}_{T}$, where $\mathcal{X}_{T}\neq\mathcal{X}_{S}(e)$ for $e = 1, \cdots , E_s.$
Consequently, we first train a sensing-aided beam prediction pre-training model in $E_s$ source domain environments and the mapping function can be expressed as 
\begin{equation}
	{f}_{\Theta}:\{\mathcal{X}(e)^{S}\}_{e=1}^{E_s} \rightarrow \{Y(e)^{S}\}_{e=1}^{E_s},
\end{equation}
where $\Theta$ is the model parameters.
\begin{algorithm}[!t]
	\caption{Transfer Learning Algorithm for Beam Prediction}
	\label{alg:4}
	\begin{algorithmic}[1]
		\REQUIRE  Source environmental feature: $\{\mathcal{X}_{S}(e)\}_{e=1}^{E_s}$, Source tasks: $\{Y(e)^{S}\}_{e=1}^{E_s}$, Target environmental feature: $\mathcal{X}_{T}$, Target task $\mathcal{Y}_{T}$
		\ENSURE   Fine-tuned parameters: $\Theta_{f}$
		\STATE \texttt{Pre-training Stage}
		\STATE \quad   ${f}_{\Theta_p}:\{\mathcal{X}(e)^{S}\}_{e=1}^{E_s} \rightarrow \{Y(e)^{S}\}_{e=1}^{E_s}$ 
		\STATE \quad Save the model parameters: $\Theta_p$
		\STATE \texttt{Fine-tuning Stage}
		\STATE \quad Scan the new static environment to obtain $\mathcal{P}_T$
		\STATE \quad Capture $\mathcal{D}_{T}$ and $\mathcal{U}_T$ during operations, 
		\STATE \quad $\mathcal{X}_T=\{\mathcal{D}_T, \mathcal{P}_T, \mathcal{U}_T\}$
		\STATE\quad  BSs capture optimal beam labels $\mathcal{Y}_{T}$ by beam scanning
		\STATE \quad ${f}_{\Theta_{\mathrm{var}},\Theta_{\mathrm{fix}}}:\mathcal{X}_T \rightarrow \mathcal{Y}_{T}$, $\Theta_f = \{\Theta_{\mathrm{var}},\Theta_{\mathrm{fix}} \}$
		\STATE \texttt{Predicting Stage}
		\STATE \quad Load fine-tuned parameters: $\Theta_f$
		\STATE \quad Capture images $\boldsymbol{X}^{(n)}$ and user location $\boldsymbol{l}^{(n)}$
		\STATE \quad $( \boldsymbol{w}_r^{opt}, \boldsymbol{w}_t^{opt} ) \leftarrow \{\boldsymbol{X}^{(\tilde{n})}, \mathcal{P}_T, \boldsymbol{l}_e^{(\tilde{n})}\}_{\tilde{n}=n}^{n+l-1}$
	\end{algorithmic}
\end{algorithm}

\begin{figure}[!t]
	\centering
	\includegraphics[width=3.5in]{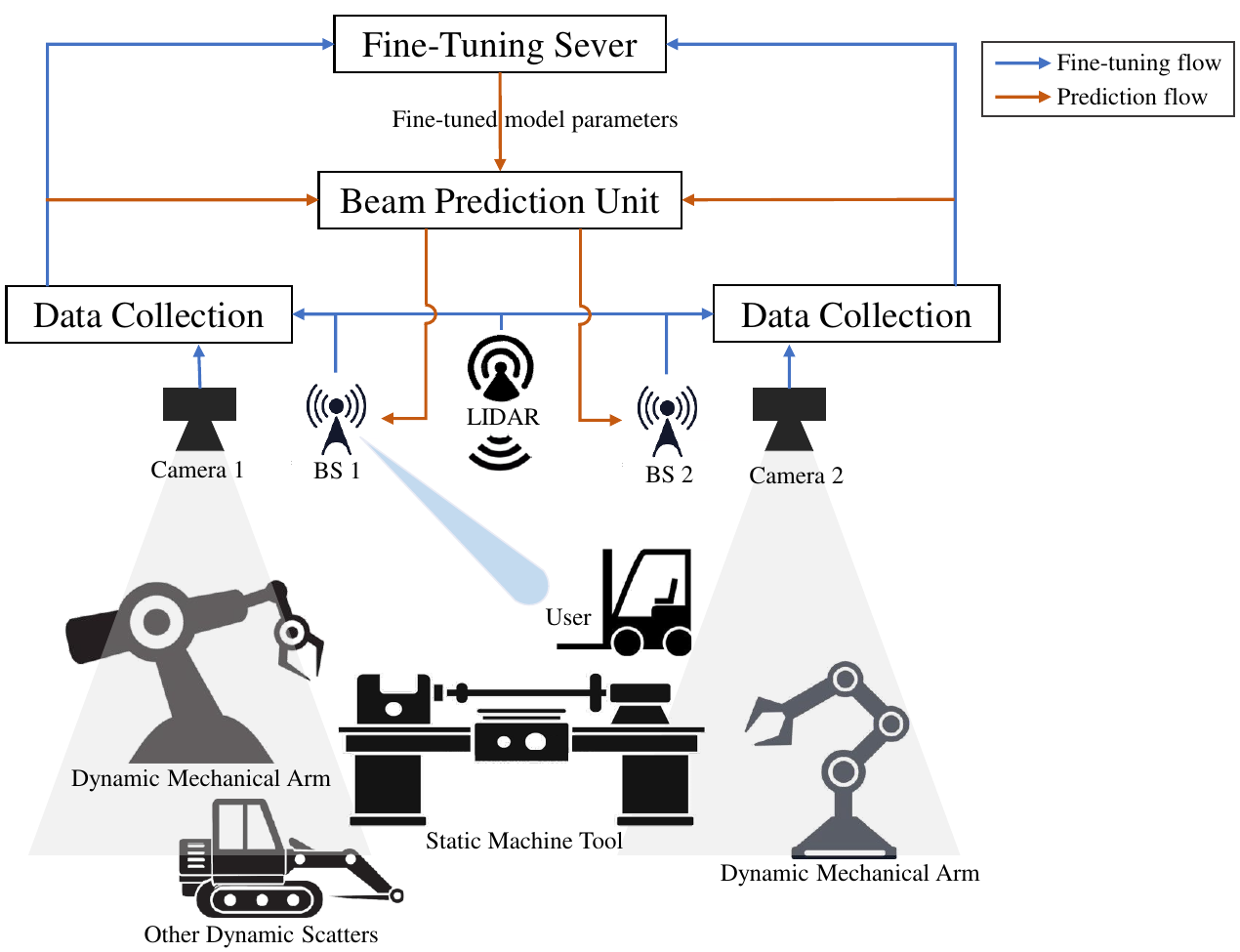}
	\caption{Diagram of the transfer learning strategy.}
	\label{fig_transfer}
\end{figure}

When facing a new environment, the fine tuning procedure will be launched as shown in Fig.~\ref{fig_transfer}. The LIADR will scan the target environment once to capture 3D static point clouds $\mathcal{P}_T$, and the 3D point clouds will be stored in the database. At the beginning stage of factory production, the cameras and the APs respectively capture environmental images and optimal beam labels $\mathcal{Y}_{T}$.
The environmental images are processed by dynamic scatterers detection to obtain $\mathcal{D}_{T}$ and controller utilizes user's locations to generate $\mathcal{U}_{T}$.
Then the $\mathcal{X}_T=\{\mathcal{D}_T, \mathcal{P}_T, \mathcal{U}_T\}$ and the $\mathcal{Y}_{T}$ are used to fine-tune the pre-trained model.
The fine-tuning mapping function can be expressed as
\begin{equation}
	{f}_{\Theta_{\mathrm{var}},\Theta_{\mathrm{fix}}}:\mathcal{X}_T \rightarrow \mathcal{Y}_{T}
\end{equation}
where $\Theta_{\mathrm{var}}$ and $\Theta_{\mathrm{fix}}$ represent the variable and fixed parameters during the fine-tuning process, respectively. Considering that the first several layers of CNNs extract low level information, and that LSTM temporal prediction is consistent across different environments, we fix the parameters of the first several layers of dynamic, static, and user identity feature extraction and the parameters of the LSTM layer. The other parameters of beam prediction model will be fine-tuned by limited labeled data.
In predicting stage, the fine-tuned parameters are loaded into the prediction model through the prediction flow. 
It is sufficient to achieve beam prediction  without pilot overhead in a new environment solely by obtaining real-time environmental images and user identities.
Note that the LIDAR scan the static environment only  once when facing a new environment. 

\section{Simulation results \label{simulation}}
In this section, we utilize simulation software to generate joint image-channel dataset for single environment and multi-environment in smart factories respectively, and evaluate the performance of the proposed environment sensing-aided beam prediction framework. 
\subsection{Dataset Generation}

\subsubsection{Image Dataset\label{image dataset}}
\begin{figure}[!t]
	\centering
	\includegraphics[width=2.8in]{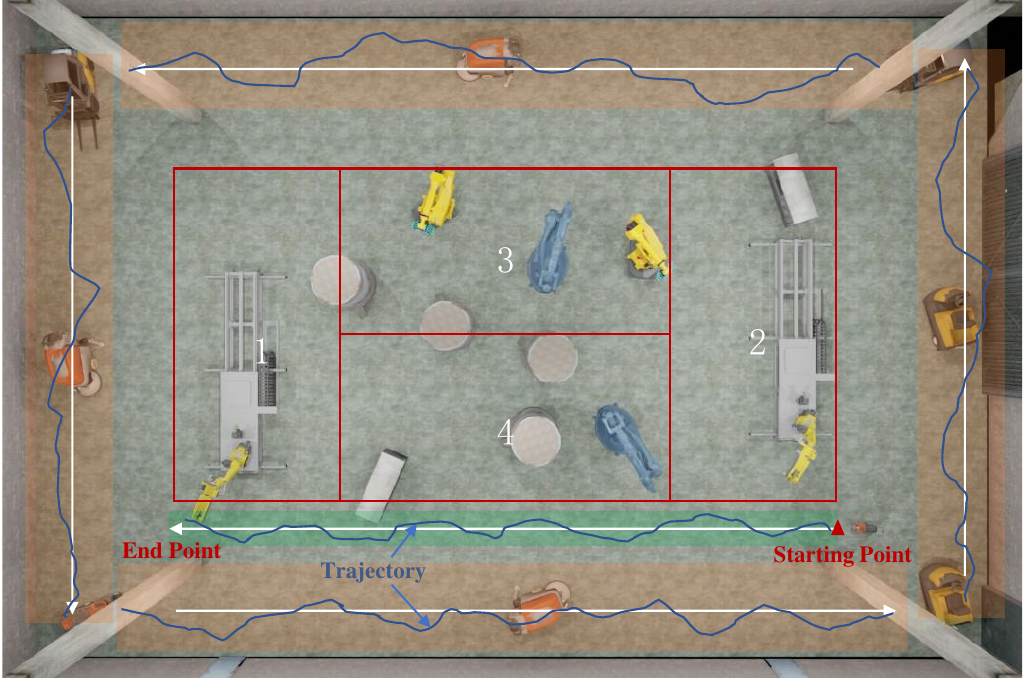}
	\caption{The top view of the smart factory.}
	\label{fig7}
\end{figure}

We utilize the Blender \cite{mullen2011mastering}, a 3D modeling software, to build a factory model with the length of  60 m $(l=60)$, width of 40 m $(l=40)$  and the height of 20 m $(h=20)$. As shown in Fig~\ref{fig7}, the area in the middle of the factory, with a length of 40 m and a width of 20 m, is the work area used to deploy equipment such as machine tools, mechanical arms, and controllers. The surrounding zone of the work area is the transportation area, where engineering vehicles can complete transportation tasks. To mimic the real world, we add randomness to the direction and speed of the engineering vehicles. Specifically, while ensuring that the engineering vehicles do not exceed the transportation area, their direction and speed follow a truncated Gaussian distribution~\cite{serfling2009approximation}. The standard form of truncated Gaussian distribution can be expressed as
\begin{equation}
	\psi(\mu, \sigma^2, a, b; x) =
	\begin{cases}
		0, & x \leq a \\
		\frac{\phi(\mu, \sigma^2; x)}{\phi(\mu, \sigma^2; b)-\phi(\mu, \sigma^2; a)}, & a<x \leq b \\
		0, &x > b
	\end{cases},
\end{equation}
where $\phi(\mu, \sigma^2; x)$ is the standard normal distribution with a mean of $\mu$ and a variance of $\sigma^2$; $a$ and $b$ are the upper bound and the lower bound of truncated Gaussian distribution respectively. If the direction of the engineering vehicle at time $t$ is $\theta_t$, then the direction at time $t+1$ can be expressed as
\begin{equation}
	\theta_{t+1} = \psi(\theta_t, 7.5, -30, 30; \theta).
\end{equation}
We take the direction of the engineering vehicle at time $t$ as the mean with 7.5 degrees as the variance, and the range is limited to $-30\sim30$ degrees to generate the direction at time $t+1$. This approach ensures a certain level of continuity in the vehicle's direction while avoiding complete randomness, which makes simulation more consistent with real world. Additionally, a similar method is applied to the speed of engineering vehicles. The speed at time $t+1$ can be expressed as
\begin{figure}[!t]
	\centering
	\subfloat[Mode 1]{\includegraphics[width=1in]{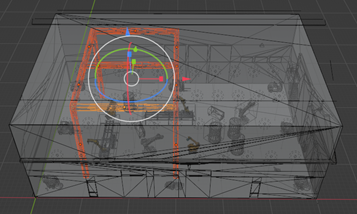}%
	}
	\hfil
	\subfloat[Mode 2]{\includegraphics[width=1in]{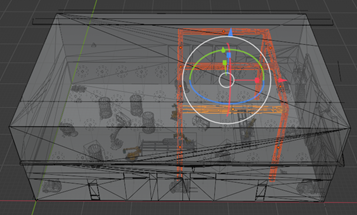}%
	}
	\hfil
	\subfloat[Mode 3]{\includegraphics[width=1in]{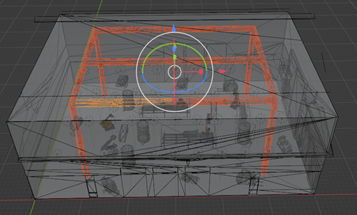}%
	}
	\caption{Three modes of the steel crossbeams and columns.}
	\label{fig10}
\end{figure}
\begin{figure}[!t]
	\centering
	\includegraphics[width=2in]{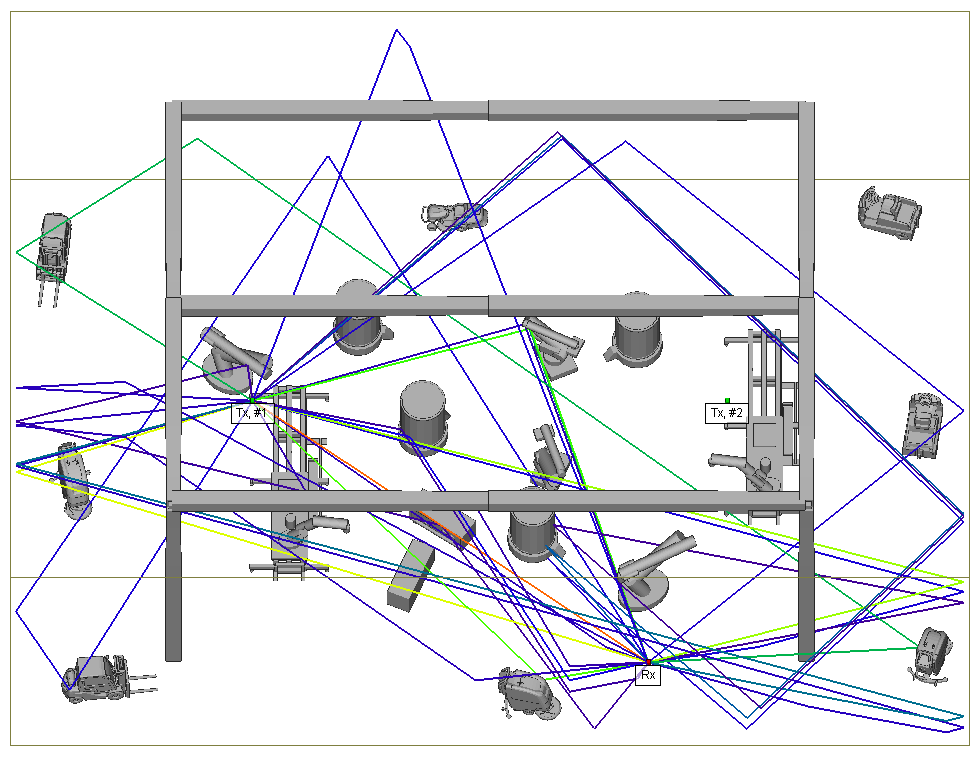}
	\caption{A visualization case of the ray tracing simulation between the BS1 and the UE.}
	\label{fig18}
\end{figure}
\begin{equation}
	v_{t+1} = \psi(v_t, 0.02, 0.8, 1.2; v).
\end{equation}
We take the speed of the engineering vehicle at time $t$ as the mean, with 0.02 m/s as the variance, and the range is limited to $0.8\sim1.2$ m/s to generate the speed at time $t+1$.
As shown in Fig.~\ref{fig7}, the user engineering vehicle starts from the right initial point and travels to the left endpoint, and its direction and speed also follow the above probability distribution.  The user's journey from the starting point to the endpoint is recorded as a \textit{path}. Due to the randomness of direction and speed, the time required for the user to complete each \textit{path} will be different.  The mechanical arms will rotate randomly at a speed of $0\sim36$ degree/s.
Then we deploy $m=2$ rows and $n=3$ columns of cameras on the ceiling of the factory to capture the environmental images at a frame rate of 20 fps and the frame interval is 50 ms. Specifically, the resolution of each image is $ W \times H$, with $W=H=360$. The six regional images will be concatenated into a global image of $1080 \times 720$. The generation schemes for single environment and multi-environment image datasets are as follows:
\begin{figure}[!t]
	\centering
	\includegraphics[width=3in]{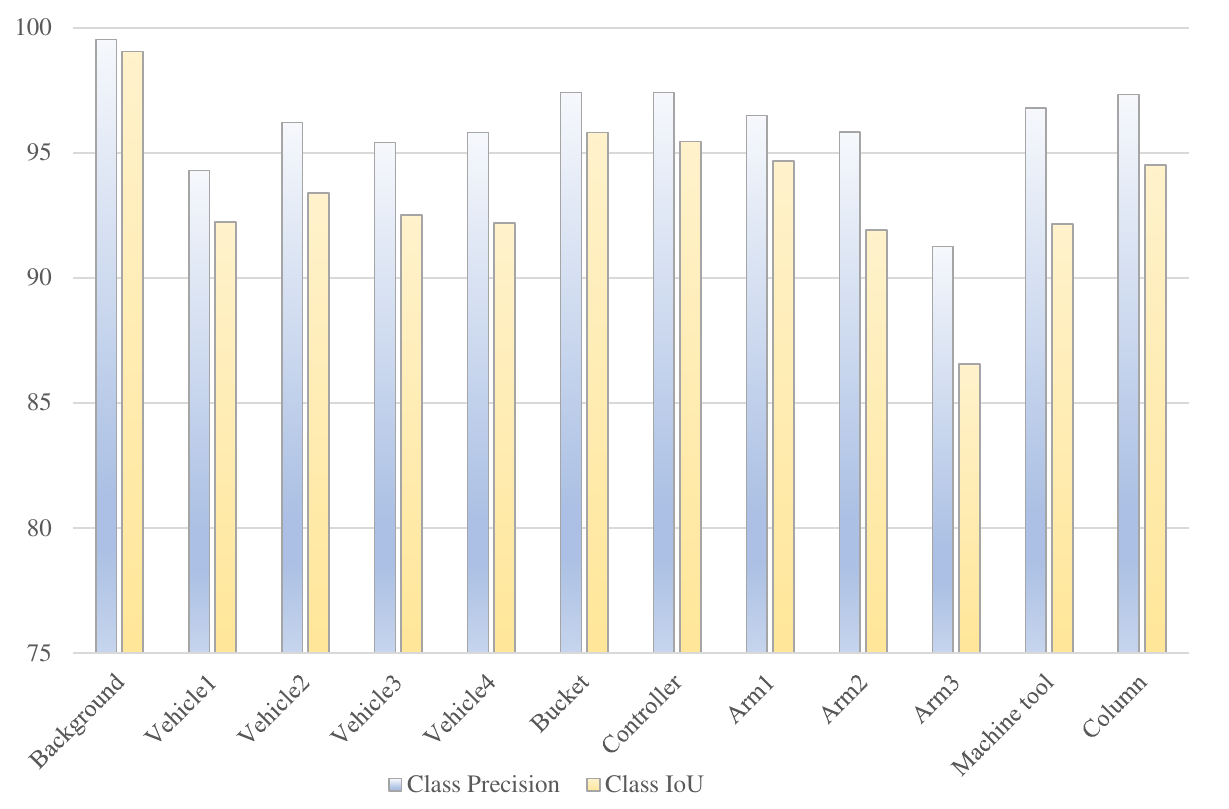}
	\caption{The segmentation precision and the intersection to union (IoU) of all semantics using Deeplabv3+ model.}
	\label{fig8}
\end{figure}
\begin{itemize}
	\item Single environment: Static equipment such as columns, machine tools, controllers, and buckets are fixed in the work area. The mechanical arms rotate randomly and the engineering vehicles travel in the transportation area. The user engineering vehicle starts from the initial point to the endpoint each time. We simulate 8 \textit{paths} with each \textit{path} ranging from 350 to 400 frames and obtain a total of 3000 images of $1080\times720$.
	\item Multi-environment: As shown in Fig.~\ref{fig7}, the work area is divided into four areas and two randomly selected areas are used to deploy two machine tools. The machine tools will be placed vertically in areas 1 and 2, and horizontally in areas 3 and 4. The location of the machine tool within the area is random and the machine tool does not exceed the boundary of the area. The mechanical arms, controllers, and other equipment are randomly deployed in the work area and do not overlap with each other. There are three modes for the steel crossbeams and columns as shown in Fig.~\ref{fig10}, and the environment will randomly choose one of the modes. The mechanical arms rotate and the engineering vehicles travel in the transportation area. The user engineering vehicle starts from the initial point to the endpoint each time. We simulate 15 \textit{paths} with each \textit{path} ranging from 350 to 400 frames and obtain a total of 5300 images of $1080\times720$.
\end{itemize}

\begin{figure*}[!t]
	\centering
	\subfloat[Segmentation map of vehicle2]{\includegraphics[width=2in]{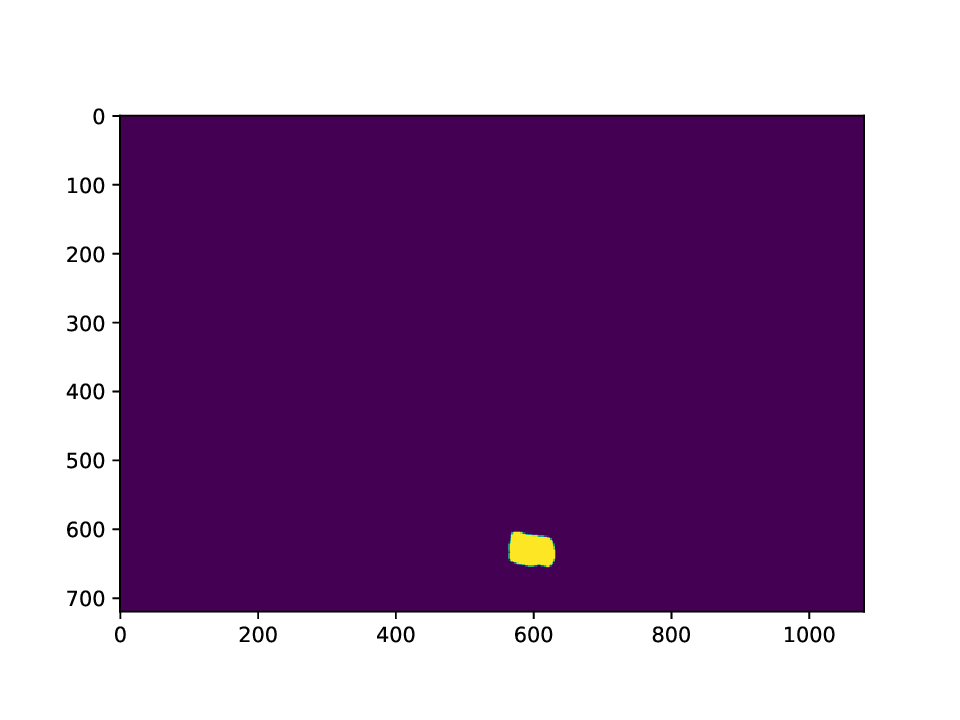}%
	}
	\hfil
	\subfloat[Segmentation map  of arm1]{\includegraphics[width=2in]{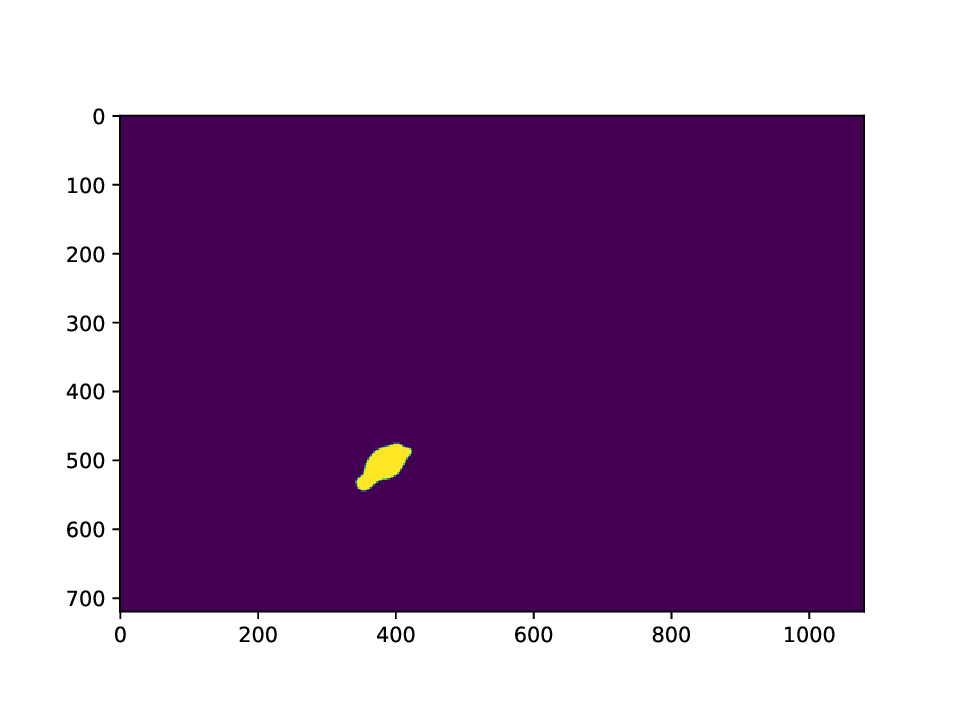}%
	}
	\hfil
	\subfloat[Segmentation map  of bucket]{\includegraphics[width=2in]{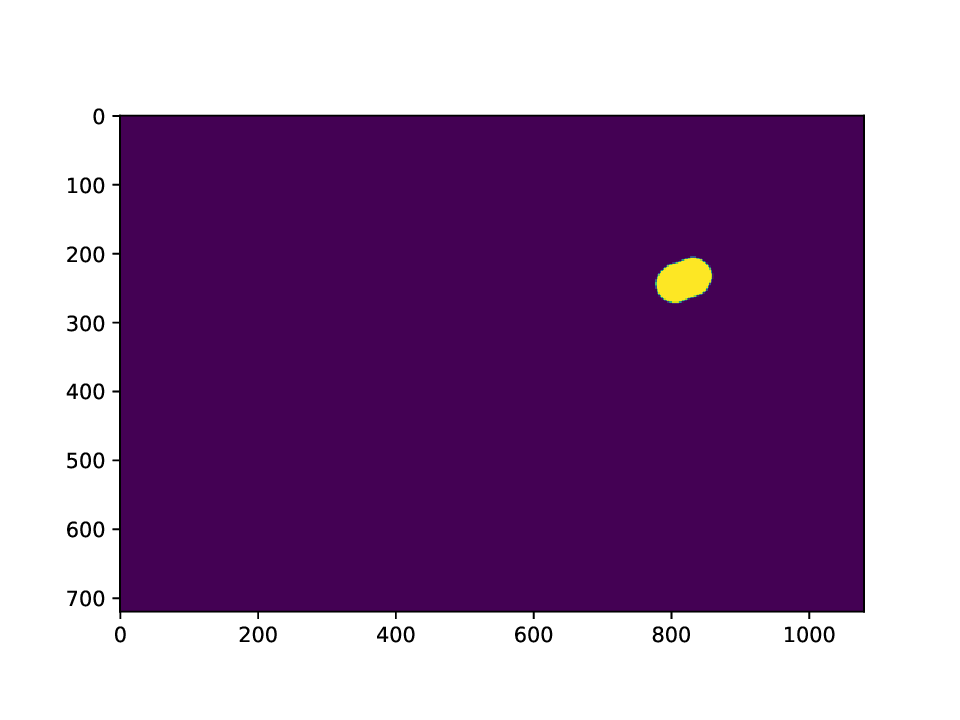}%
	}
	\hfil
	\subfloat[Pixel number difference sequence $\boldsymbol{s}_{q,d}^{(n)}$ of vehicle2 ]{\includegraphics[width=2in]{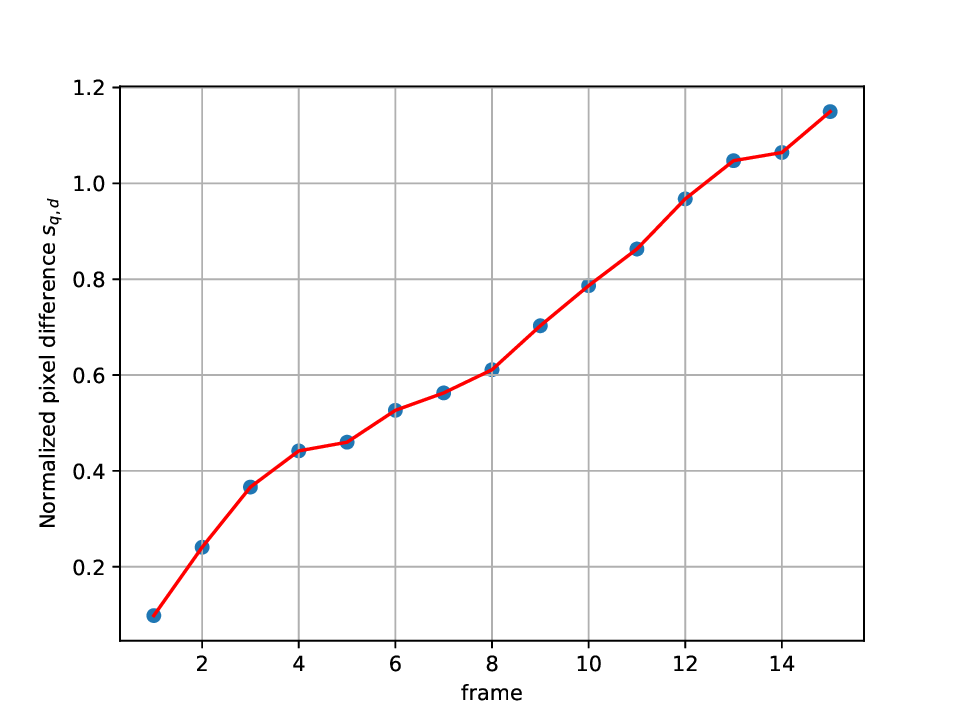}%
	}
	\hfil
	\subfloat[Pixel number difference sequence $\boldsymbol{s}_{q,d}^{(n)}$ of arm1]{\includegraphics[width=2in]{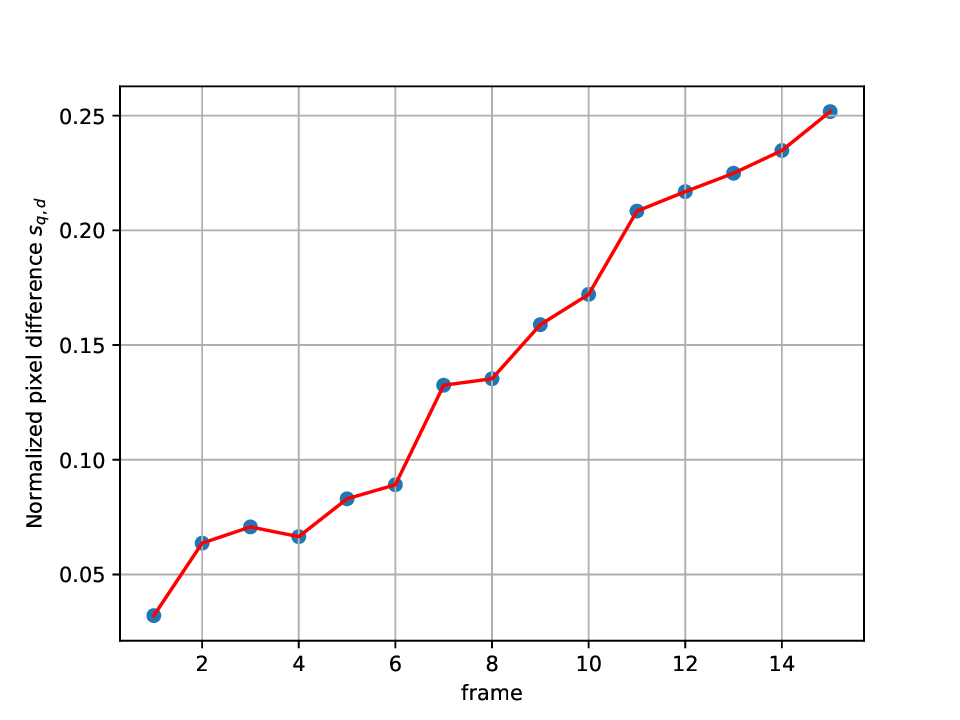}%
	}
	\hfil
	\subfloat[Pixel number difference sequence $\boldsymbol{s}_{q,d}^{(n)}$ of bucket]{\includegraphics[width=2in]{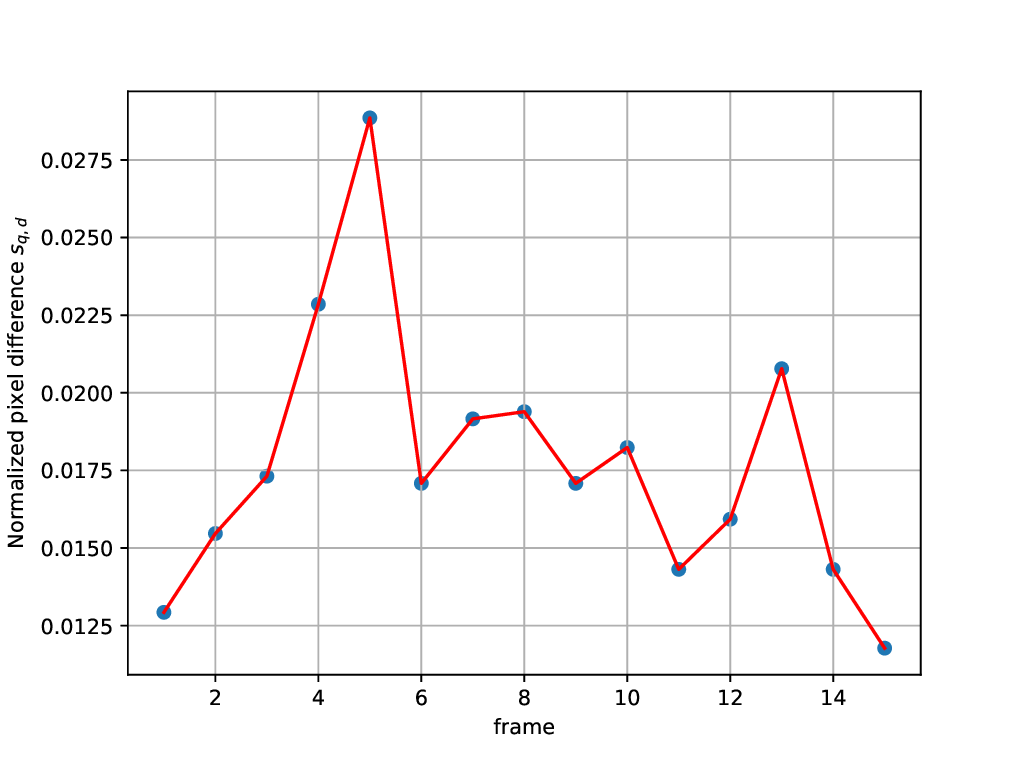}%
	}
	\hfil
	\caption{Segmentation maps  of several instances and their difference sequence  $\boldsymbol{s}_{q,d}^{(n)}$ versus frame. The difference sequence of dynamic scatterers shows a continuous increasing trend, while the difference sequence of the stationary bucket shows an irregular fluctuation which is similar to white noise.}
	\label{fig9}
\end{figure*}

\subsubsection{Channel Dataset}
The channel dataset is generated from Wireless Insite~\cite{WinNT}, a simulation software for radio signals and wireless communications systems using ray tracing model~\cite{glassner1989introduction}. We first import the 3D model of the building and scatterers into Wireless Insite frame by frame. Then we set the material of factory buildings as concrete, while set the material of equipment as metal. There are $B=2$ transmitters with fixed locations and their coordinates are (6, 30, 8) and (30, 6, 8) as shown in Fig.~\ref{fig2}. The receiver is deployed on top of the user engineering vehicle. Next, we set the simulation parameters of the  wireless system including signal frequency as 28 GHz, waveform as sinusoid, antennas as half-wave dipole, polarization as vertical, voltage standing wave ratio (VSWR) as 1.0, and power as 0 dBm. We adopt the X3D ray model~\cite{X3D} to simulate the parameters of the ray paths. A visualization of the ray tracing simulation between BS1 and the UE is shown in Fig.~\ref{fig18}. We select the parameters of the top 25 paths with the strongest power including received signal power $P_r$ (dBm), received signal phase $\Phi$ (deg), azimuth AOA $\theta_r $ (deg), elevation AOA $\phi_r$ (deg), azimuth AOD $\theta_t$ (deg) and elevation AOD $\phi_t$ (deg). Then we use these simulated parameters to calculate the channel matrix ${\boldsymbol{H}}_{k, i}$ based on Eq.~(\ref{equ2}).

\subsection{Performance of the Semantics Segmentation and Dynamic Semantics Separation}

We adopt Deeplabv3+ model to segment environmental images. As shown in Fig.~\ref{fig8}, the segmentation precision and the intersection to union (IoU) of all classes has reached over 90\% and 85\% respectively. The mean precision $	A_{\mathrm{Dpv3+}}$ and mean intersection to union $\mathrm{mIOU}$ are 96.15\% and 93.37\% respectively. High precision and mIoU will provide support and guarantee for subsequent dynamic semantic separation.
Fig.~\ref{fig9} exhibits the semantics segmentation and the difference sequence $\boldsymbol{s}_{q,d}^{(n)}$ versus frame. We can find that due to the movement and the rotation of dynamic scatterers, their $\boldsymbol{s}_{q,d}^{(n)}$ shows a continuous increasing trend. The difference sequence $\boldsymbol{s}_{q,d}^{(n)}$ of the stationary bucket shows an irregular fluctuation due to semantic segmentation error, which is similar to white noise. Therefore, we utilize Ljung-Box test to determine the autocorrelation of the sequence $\boldsymbol{s}_{q,d}^{(n)}$, which can retain the dynamic scatterers and filter out the static environment. Specifically, we compute different pixel values number of continuous $l=16$ frames, with $\mathrm{dim}(\boldsymbol{s}_{q,d}^{(n)})=15$. We set the lag order of Ljung Box test $m=4$ and the significance level $\alpha=0.15$. The decision rule is
\begin{equation}
	{H_0} =	\left\{
	\begin{aligned}
		{\rm received} &,\quad Q(4)\leq \chi^2_{0.85}(4)\\
		{\rm rejected} &,\quad Q(4)>\chi^2_{0.85}(4).
	\end{aligned}
	\right.
\end{equation}
When $H_0$ is received, the instance will be judged as static environment and is filtered out. When $H_0$ is rejected, the instance will be determined as a dynamic scatterer and is retained. 
After testing, the accuracy of dynamic scatterers separation reaches 95\%. The reason for confusing dynamic scatterers and static environment is semantic segmentation errors. When there is a significant deviation in the semantic pixels of a certain frame or several frames, the difference sequece $\boldsymbol{s}_{q,d}^{(n)}$ will be affected, leading to the judgment errors. If semantic segmentation is completely accurate, then the accuracy of dynamic scatterers separation will be 100\%.

\subsection{Performance of the Singe environment}
\begin{figure}[!t]
	\centering
	\includegraphics[width=3in]{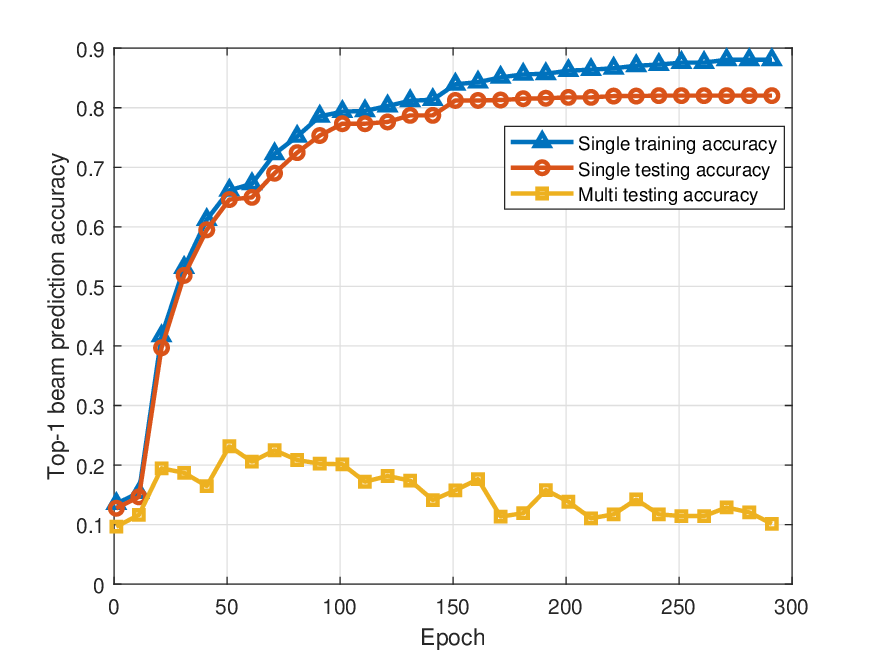}
	\caption{The Top-1 beam prediction accuracy 400 ms in advance on single environment training set, single environment testing set, and multi-environment testing set versus the number of training epochs.}
	\label{fig11}
\end{figure}

To demonstrate that a model trained in a single environment cannot be directly applied to multiple environments, we divide the single environment data into training set (2400 samples) and testing set (600 samples) in an 8:2 ratio.
Then, we randomly select 3 \textit{path} data (1050 samples) out of 15 \textit{paths} in multiple environments as the multi-environment testing set. The Top-1 beam prediction accuracy 400 ms in advance versus the number of training epochs is shown in Fig.~\ref{fig11}. It can be seen that as the training set accuracy of single environment  improves, the testing set accuracy of single environment  also continues to rise, while the testing set accuracy of the multi-environment remains around 0.1. The reason is that although dynamic features, static features, and user identity features are already sufficient to characterize environment and channel information, for data-driven NNs, a single environment is only one sample in all environment sample spaces. The mapping model that only learn from a single environment often overfits this single environment and do not have the ability to generalize to other environments. However, the variable space of the multiple environments is very large, and in practical applications, it is not possible to collect all potential environment data. Therefore, we propose to use fine-tuning techniques in transfer learning to generalize the model  from the source domain environments to a target domain environment, thereby reducing the cost of data collection and training time.
\subsection{Performance of the Multi-environment}
\begin{figure}[!t]
	\centering
	\includegraphics[width=3in]{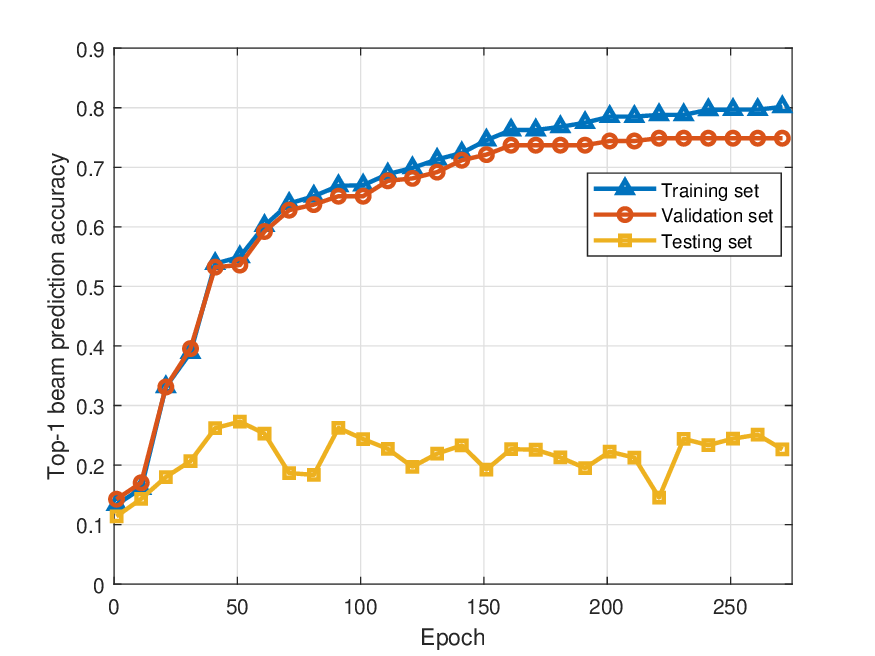}
	\caption{The Top-1 beam prediction accuracy 400 ms in advance on multi-environment training set, validation set, and testing set versus the number of training epochs.}
	\label{fig12}
\end{figure}
\begin{figure}[!t]
	\centering
	\includegraphics[width=3in]{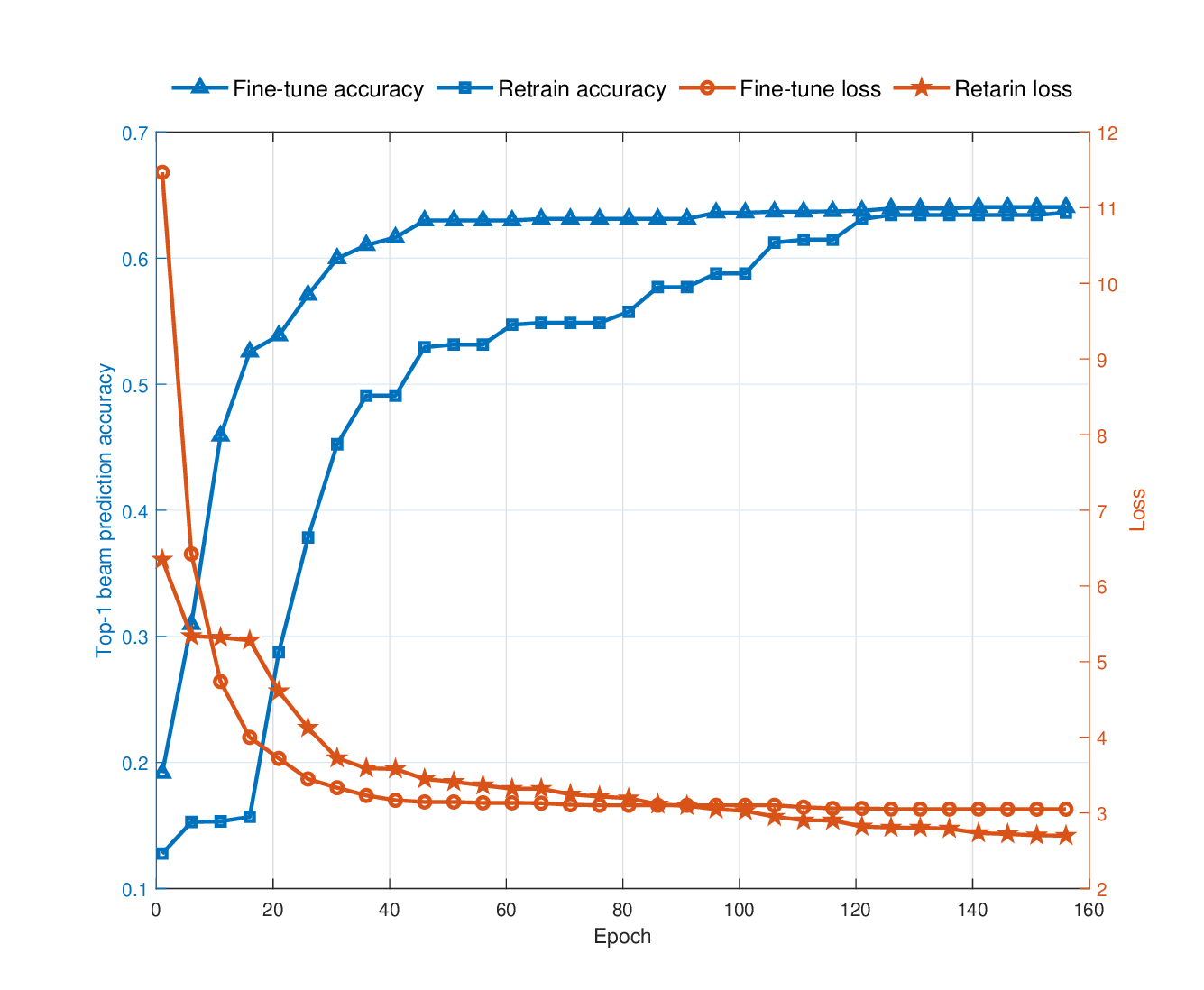}
	\caption{The Top-1 accuracy and the loss of the pre-trained model and the untrained model on testing set versus the number of training epochs.}
	\label{fig13}
\end{figure}
We take the multi-environment data as the source domain data and the single environment data as the target domain data. Then we divide the 15 \textit{path} data (5300 samples) from the multi-environment dataset into a training set (4240 samples) and a validation set (1060 samples) in an 8:2 ratio, and 30\% of the single environment data (900 samples) are taken as the testing set. Fig.~\ref{fig12} displays the Top-1 beam prediction accuracy 400~ms in advance on training set, validation set, and testing set versus the number of training epochs. It can be seen that the accuracy of the training set and validation set increase synchronously, while the accuracy of the testing set is around 0.2$\sim$0.3. 
This indicates that the model trained in a limited number of environments has better generalization performance than the model trained in one environment, but still cannot be transferred to a new environment.
The reason is that the sample space of the environment is large, and the mapping model learned from only a limited number of environments often overfits these environments and do not have the ability to generalize to other environments.
\begin{figure}[!t]
	\centering
	\subfloat[]{\includegraphics[width=1.7in]{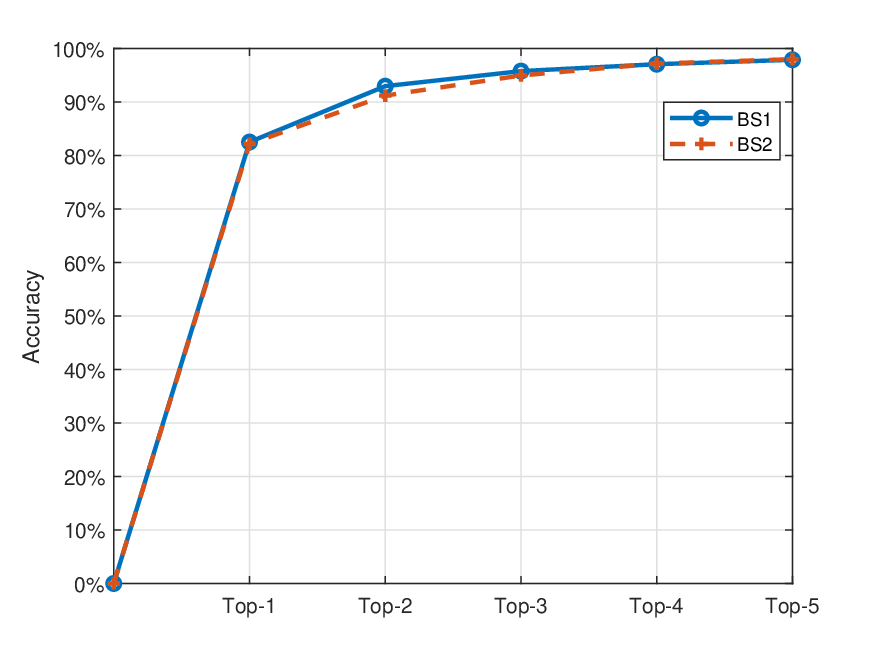}%
	}
	\hfil
	\subfloat[]{\includegraphics[width=1.7in]{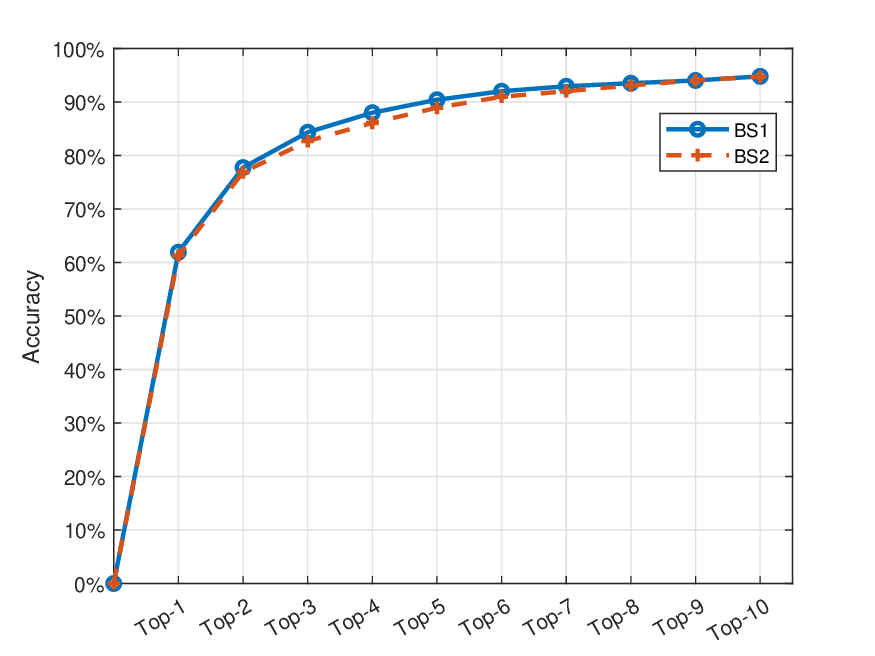}%
	}
	\caption{The Top-5 beam prediction accuracy on multi-environment validation set and the fine-tuned Top-10 beam prediction accuracy on testing set.}
	\label{fig14}
\end{figure}
However, there are commonalities in feature extraction for different environments, and the factors that affect channel variations are also consistent. Therefore, a part of the pre-trained model parameters for a limited number of environments can be fixed and a small portion of data from the target environment can be used to fine tune the model to achieve environment generalization. Specifically, we save the model parameters with the highest accuracy in the validation set, with a Top-1 accuracy of 82\%. Then, we fix the parameters of the first two layers of feature extraction and the parameters of the LSTM layer. The 30\% of the single environment data (900 samples) are selected to fine tune the pre-trained model, while the remaining 70\% are used as the testing set. Moreover, as a comparison, we retrain a beam prediction model using 30\% of the single environment data.
The accuracy and the loss of the pre-trained model and the untrained model on testing set versus the number of training epochs are shown in Fig.~\ref{fig13}. It can be seen that the Top-1 accuracy of the pre-trained model rapidly increases to 61\% at the $40$-th epoch, while the Top-1 accuracy of the untrained model did not reach the same level until the $160$-th epoch, which indicates that there is indeed a certain similarity between the source environments and the target environment and pre-training can save 75\% of the training time. After fine-tuned by 30\% data from target environment, the pre-trained model can be well transferred to the a new environment. Fig.~\ref{fig14} displays the Top-5 accuracy on the validation set and the fine-tuned Top-10 accuracy on the testing set. The Top-5 accuracy of the validation set reaches 97\%, while the Top-5 accuracy of fine-tuned model in target environment is 90\%, but its Top-10 accuracy still reaches 94\%. Although the performance of the fine-tuned model in the target environment has decreased compared to the original model in the source environments, the proportion of decline is relatively low, and the accuracy can still meet practical needs. The proposed scheme transfers the environment sensing-aided beam prediction model from given environments to a new environment while only collects 30\% of labeled data from the new environment.

\section{Conclusion \label{conclusion}}
In this paper, we have proposed an environment sensing-aided beam prediction model that can be transferred from given environments to a new environment. Simulation results show that the Top-5 beam prediction of the pre-training model reaches 97\% in source environments, and the Top-10 beam prediction accuracy of the fine-tuned model reaches 94\% in target environment. Moreover, the amount of training data and the cost of training time save 70\% and 75\% respectively.
The proposed transfer learning strategy enable sensing-aided beam prediction model to rapidly adapt to a new environment, which greatly saves the cost of data collection and training time, and enhances the ability of deep model to adapt to real-world variations.

\bibliographystyle{IEEEtran}
\linespread{0.9}
\bibliography{reference.bib}

\begin{thebibliography}{10}
\providecommand{\url}[1]{#1}
\csname url@samestyle\endcsname
\providecommand{\newblock}{\relax}
\providecommand{\bibinfo}[2]{#2}
\providecommand{\BIBentrySTDinterwordspacing}{\spaceskip=0pt\relax}
\providecommand{\BIBentryALTinterwordstretchfactor}{4}
\providecommand{\BIBentryALTinterwordspacing}{\spaceskip=\fontdimen2\font plus
\BIBentryALTinterwordstretchfactor\fontdimen3\font minus
  \fontdimen4\font\relax}
\providecommand{\BIBforeignlanguage}[2]{{%
\expandafter\ifx\csname l@#1\endcsname\relax
\typeout{** WARNING: IEEEtran.bst: No hyphenation pattern has been}%
\typeout{** loaded for the language `#1'. Using the pattern for}%
\typeout{** the default language instead.}%
\else
\language=\csname l@#1\endcsname
\fi
#2}}
\providecommand{\BIBdecl}{\relax}
\BIBdecl

\bibitem{s22072688}
\BIBentryALTinterwordspacing
A.~Jabbar, Q.~H. Abbasi, N.~Anjum, T.~Kalsoom, N.~Ramzan, S.~Ahmed, P.~M.
  Rafi-ul Shan, O.~P. Falade, M.~A. Imran, and M.~Ur~Rehman, ``Millimeter-wave
  smart antenna solutions for {URLLC} in {I}ndustry 4.0 and beyond,''
  \emph{Sensors}, vol.~22, no.~7, Mar. 2022. [Online]. Available:
  \url{https://www.mdpi.com/1424-8220/22/7/2688}
\BIBentrySTDinterwordspacing

\bibitem{meindl2021four}
B.~Meindl, N.~F. Ayala, J.~Mendon{\c{c}}a, and A.~G. Frank, ``The four smarts
  of {I}ndustry 4.0: Evolution of ten years of research and future
  perspectives,'' \emph{Technol. Forecast. Soc. Chang.}, vol. 168, pp. 120,784,
  Jul. 2021.

\bibitem{lasi2014industry}
H.~Lasi, P.~Fettke, H.-G. Kemper, T.~Feld, and M.~Hoffmann, ``Industry 4.0,''
  \emph{Business \& {I}nf. {S}yst. {E}ng.}, vol.~6, no.~4, pp. 239--242, Jun.
  2014.

\bibitem{9145564}
I.~F. Akyildiz, A.~Kak, and S.~Nie, ``6{G} and beyond: {T}he future of wireless
  communications systems,'' \emph{IEEE Access}, vol.~8, pp. 133\,995--134\,030,
  Jul. 2020.

\bibitem{9390169}
H.~Tataria, M.~Shafi, A.~F. Molisch, M.~Dohler, H.~Sjöland, and F.~Tufvesson,
  ``6{G} wireless systems: {V}ision, requirements, challenges, insights, and
  opportunities,'' \emph{Proc. IEEE}, vol. 109, no.~7, pp. 1166--1199, Mar.
  2021.

\bibitem{LU2020100158}
\BIBentryALTinterwordspacing
Y.~Lu and X.~Zheng, ``6{G}: {A} survey on technologies, scenarios, challenges,
  and the related issues,'' \emph{J. Ind. Inf. Integr.}, vol.~19, p. 100158,
  Sep. 2020. [Online]. Available:
  \url{https://www.sciencedirect.com/science/article/pii/S2452414X20300339}
\BIBentrySTDinterwordspacing

\bibitem{andersen1995propagation}
J.~B. Andersen, T.~S. Rappaport, and S.~Yoshida, ``Propagation measurements and
  models for wireless communications channels,'' \emph{IEEE Commun. Mag.},
  vol.~33, no.~1, pp. 42--49, Jan. 1995.

\bibitem{9529181}
F.~Gao, B.~Lin, C.~Bian, T.~Zhou, J.~Qian, and H.~Wang, ``Fusionnet: {E}nhanced
  beam prediction for mmwave communications using sub-6 {GHz} channel and a few
  pilots,'' \emph{IEEE Trans. Commun.}, vol.~69, no.~12, pp. 8488--8500, Sep.
  2021.

\bibitem{9771564}
U.~Demirhan and A.~Alkhateeb, ``Radar aided 6{G} beam prediction: {D}eep
  learning algorithms and real-world demonstration,'' in \emph{Proc. IEEE
  Wireless Commun. Netw. Conf. (WCNC)}, Apr. 2022, pp. 2655--2660.

\bibitem{8642397}
A.~Klautau, N.~González-Prelcic, and R.~W. Heath, ``{LIDAR} data for deep
  learning-based mmwave beam-selection,'' \emph{IEEE Wireless Commun. Lett.},
  vol.~8, no.~3, pp. 909--912, Feb. 2019.

\bibitem{marasinghe2021lidar}
D.~Marasinghe, N.~Rajatheva, and M.~Latva-aho, ``{LiDAR} aided human blockage
  prediction for 6{G},'' in \emph{Proc. IEEE Globecom Workshops (GC Wkshps)},
  Dec. 2021, pp. 1--6.

\bibitem{10412143}
Y.~Feng, F.~Gao, X.~Tao, S.~Ma, and H.~V. Poor, ``Vision-aided ultra-reliable
  low-latency communications for smart factory,'' \emph{IEEE Trans. Commun.},
  pp. 1--1, Jan. 2024.

\bibitem{alrabeiah2020millimeter}
M.~Alrabeiah, A.~Hredzak, and A.~Alkhateeb, ``Millimeter wave base stations
  with cameras: {V}ision-aided beam and blockage prediction,'' in \emph{Proc.
  IEEE 91st Veh. Technol. Conf. (VTC)}, May. 2020, pp. 1--5.

\bibitem{xu2022computer}
W.~Xu, F.~Gao, X.~Tao, J.~Zhang, and A.~Alkhateeb, ``Computer vision aided
  mm{W}ave beam alignment in {V2X} communications,'' \emph{IEEE Trans. Wireless
  Commun.}, vol.~22, no.~4, pp. 2699--2714, Oct. 2022.

\bibitem{chen2022computer}
J.~Chen, F.~Gao, X.~Tao, G.~Liu, C.~Pan, and A.~Alkhateeb, ``Computer vision
  aided codebook design for {MIMO} communications systems,'' \emph{IEEE Trans.
  Wireless Commun.}, Nov. 2022.

\bibitem{nishio2019proactive}
T.~Nishio, H.~Okamoto, K.~Nakashima, Y.~Koda, K.~Yamamoto, M.~Morikura,
  Y.~Asai, and R.~Miyatake, ``Proactive received power prediction using machine
  learning and depth images for mm{W}ave networks,'' \emph{IEEE J. Select.
  Areas Commun.}, vol.~37, no.~11, pp. 2413--2427, Aug. 2019.

\bibitem{shi2015convolutional}
X.~Shi, Z.~Chen, H.~Wang, D.-Y. Yeung, W.-K. Wong, and W.-c. Woo,
  ``Convolutional {LSTM} network: {A} machine learning approach for
  precipitation nowcasting,'' \emph{Advances Neural Inform. Processing Syst.
  (NIPS)}, vol.~28, Dec. 2015.

\bibitem{charan2021vision}
G.~Charan, M.~Alrabeiah, and A.~Alkhateeb, ``Vision-aided 6{G} wireless
  communications: {B}lockage prediction and proactive handoff,'' \emph{IEEE
  Trans. Veh. Technol.}, vol.~70, no.~10, pp. 10\,193--10\,208, Aug. 2021.

\bibitem{yang2023environment}
Y.~Yang, F.~Gao, X.~Tao, G.~Liu, and C.~Pan, ``Environment semantics aided
  wireless communications: {A} case study of mm{W}ave beam prediction and
  blockage prediction,'' \emph{IEEE J. Select. Areas Commun.}, May. 2023.

\bibitem{cheffena2016industrial}
M.~Cheffena, ``Industrial wireless communications over the millimeter wave
  spectrum: {Opportunities} and challenges,'' \emph{IEEE Commun. Mag.},
  vol.~54, no.~9, pp. 66--72, Sep. 2016.

\bibitem{ali2017millimeter}
A.~Ali, N.~Gonz{\'a}lez-Prelcic, and R.~W. Heath, ``Millimeter wave
  beam-selection using out-of-band spatial information,'' \emph{IEEE Trans.
  Wireless Commun.}, vol.~17, no.~2, pp. 1038--1052, Nov. 2017.

\bibitem{lateef2019survey}
F.~Lateef and Y.~Ruichek, ``Survey on semantic segmentation using deep learning
  techniques,'' \emph{Neurocomputing}, vol. 338, pp. 321--348, Apr. 2019.

\bibitem{chen2018encoder}
L.-C. Chen, Y.~Zhu, G.~Papandreou, F.~Schroff, and H.~Adam, ``Encoder-decoder
  with atrous separable convolution for semantic image segmentation,'' in
  \emph{Proc. European Conf. Comput. Vis. (ECCV)}, Sep. 2018, pp. 801--818.

\bibitem{amari1993backpropagation}
S.-i. Amari, ``Backpropagation and stochastic gradient descent method,''
  \emph{Neurocomputing}, vol.~5, no. 4-5, pp. 185--196, Jun. 1993.

\bibitem{wu2009optimizing}
K.~Wu, E.~Otoo, and K.~Suzuki, ``Optimizing two-pass connected-component
  labeling algorithms,'' \emph{Pattern Anal. Appl.}, vol.~12, pp. 117--135,
  Mar. 2008.

\bibitem{ljung1978measure}
G.~M. Ljung and G.~E. Box, ``On a measure of lack of fit in time series
  models,'' \emph{Biometrika}, vol.~65, no.~2, pp. 297--303, Aug. 1978.

\bibitem{stationary}
\BIBentryALTinterwordspacing
H.~Cramer, ``On the theory of stationary random processes,'' \emph{Annals of
  Mathematics}, vol.~41, no.~1, pp. 215--230, Jan. 1940. [Online]. Available:
  \url{http://www.jstor.org/stable/1968827}
\BIBentrySTDinterwordspacing

\bibitem{lancaster2005chi}
H.~O. Lancaster and E.~Seneta, ``Chi-square distribution,'' \emph{Encyclopedia
  of biostatistics}, vol.~2, Jul. 2005.

\bibitem{he2016deep}
K.~He, X.~Zhang, S.~Ren, and J.~Sun, ``Deep residual learning for image
  recognition,'' in \emph{Proc. IEEE Conf. Comput. Vis. Pattern Recog. (CVPR)},
  Jun. 2016, pp. 770--778.

\bibitem{maturana2015voxnet}
D.~Maturana and S.~Scherer, ``Voxnet: A 3{D} convolutional neural network for
  real-time object recognition,'' in \emph{Proc. IEEE/RSJ Int. Conf. Intell.
  robots Syst. (IROS)}, Sep. 2015, pp. 922--928.

\bibitem{hochreiter1997long}
S.~Hochreiter and J.~Schmidhuber, ``Long short-term memory,'' \emph{Neural
  Comput.}, vol.~9, no.~8, pp. 1735--1780, Nov. 1997.

\bibitem{mullen2011mastering}
T.~Mullen, \emph{Mastering {B}lender}.\hskip 1em plus 0.5em minus 0.4em\relax
  Hoboken, NJ, USA: Wiley, 2011.

\bibitem{serfling2009approximation}
R.~J. Serfling, \emph{Approximation Theorems of Mathematical Statistics}.\hskip
  1em plus 0.5em minus 0.4em\relax John Wiley \& Sons, 2009.

\bibitem{WinNT}
\BIBentryALTinterwordspacing
Remcom. {``Wireless InSite"}. [Online]. Available:
  \url{https://www.remcom.com/wireless-insite-em-propagation-software}
\BIBentrySTDinterwordspacing

\bibitem{glassner1989introduction}
A.~S. Glassner, \emph{An {I}ntroduction to {R}ay {T}racing}.\hskip 1em plus
  0.5em minus 0.4em\relax Amsterdam, The Netherlands: Elsevier, 1989.

\bibitem{X3D}
\BIBentryALTinterwordspacing
Remcom. {``X3D"}. [Online]. Available:
  \url{https://www.remcom.com/wireless-insite-models/high-fidelity-ray-tracing}
\BIBentrySTDinterwordspacing

\end{thebibliography}

\end{document}